\documentclass{aa}

\usepackage{amsmath}
\usepackage{amssymb}
\usepackage{mathtools}
\usepackage{graphicx}
\usepackage{lscape}
\usepackage{booktabs}
\usepackage{color}
\usepackage{subfigure}
\usepackage[version=4]{mhchem}
\usepackage{isotope}
\usepackage{txfonts}

\usepackage{natbib}
\bibpunct{(}{)}{;}{a}{}{,}

\graphicspath{{./}{Figures_pdf/}}

\usepackage{hyperref}
\hypersetup{
    colorlinks=true,
    citecolor=blue,
    linkcolor=magenta,
    filecolor=magenta
    }
    
\begin{document} 

   \titlerunning{Neutrino-driven explosive nucleosynthesis}
   \authorrunning{Boccioli \& Roberti}
   \title{Explodability matters: How neutrino-driven explosions change explosive nucleosynthesis yields}
  
   \author{L. Boccioli\thanks{lbocciol@berkeley.edu}
          \inst{1}
          \and
          L. Roberti 
          \inst{2}
          }

   \institute{Department of Physics, University of California, Berkeley, CA 94720, USA\\
              \email{lbocciol@berkeley.edu}
    \and Istituto Nazionale di Fisica Nucleare - Laboratori Nazionali del Sud, Via Santa Sofia 62, Catania, I-95123, Italy\\ \email{roberti@lns.infn.it}
             }

   \date{Received October 16, 2025; accepted March 28, 2026}

  \abstract
   {Explosive nucleosynthesis is affected by many uncertainties, particularly regarding the assumptions and prescriptions adopted during a star's evolution. Moreover, simple explosion models are often used, which can introduce large errors in the assumed explosion energy and mass cut.
   In this paper, our goal is to analyze the explosion properties and nucleosynthesis of a wide range of progenitors from three different stellar evolution codes: FRANEC, KEPLER, and MESA. In particular, we show the differences between the neutrino-driven explosions simulated in this work and the much simpler bomb and piston models typically used in the literature. We then focus on the impact of different explodabilities and different explosion dynamics on the nucleosynthetic yields. 
   We adopted the neutrino-driven core-collapse supernova explosion code GR1D+, i.e., a spherically symmetric model with state-of-the-art microphysics and neutrino transport and a time-dependent mixing-length model for neutrino-driven convection. We carried out explosions up to several seconds after bounce, then calculated the nucleosynthetic yields with the post-processing code SkyNet.
   We find that our 1D+ simulations yield explosion energies and remnant masses in agreement with observations of type II-P, IIb, and Ib supernovae, as well as with the most recent 3D simulations of the explosion. We provide a complete set of yields for all the stars simulated, including rotating, low-metallicity, and binary progenitors. Finally, we find that piston and bomb models, compared with more realistic neutrino-driven explosions, can artificially increase the production of Fe-peak elements, whereas the different explodabilities cause discrepancies in lighter elements.}
  
   \keywords{Physical data and processes: Nuclear reactions, nucleosynthesis, abundances --
             Stars: massive --
             Stars: abundances --
             Stars: supernovae: general --
             ISM: abundances}

   \maketitle
    
\section{Introduction} 
\label{sec:intro}

Core-collapse supernovae (CCSNe) are responsible for enriching the interstellar medium (ISM) with almost half of the elements in the periodic table \citep{BBFH_1957,Arnett1987,Kobayashi2020_OriginElements}. Broadly speaking, these elements can be synthesized either in the hydrostatic burning processes during the post-main-sequence phase of the star \citep{Weaver1978_presn_evolution,Aufderheide1991_bomb,Limongi2003_Evolution_Expl_Nuc,Nomoto2013_Nucleosynthesis_in_Stars,Hix1999_XNet_methods} and then ejected by the supernova, or during the explosion itself \citep{Arnett1969_ExplNuc,Limongi2003_Evolution_Expl_Nuc,Arcones2023_OriginOfElements,Frohlich2006_CFNET_methods}. 

Both the pre-supernova and explosive nucleosynthesis yields are swept up by the supersonic bubble that propagates into the ISM and can therefore be visible in the ensuing supernova light curve \citep{Gal-Yam2017_SN_classification} as well as, at later times, in the supernova remnant \citep{Vink2020_SN_remnants_review,Milisavljevic2024_JWST_CasA}. Eventually, they completely mix into the ISM and make up the material from which the next generation of stars form \citep{Diehl2020_GCE_living_review,Kobayashi2020_OriginElements}. Therefore, accurate predictions of both the pre-supernova and the explosive yields are crucial for explaining the observed supernova light curves and spectra, the observations of supernova remnants, and the derived photospheric spectral abundances of stars.

The uncertainties that dominate current predictions of CCSN yields are due both to nuclear physics (i.e., uncertainties in reaction rates) and astrophysics (i.e., uncertainties in the astrophysical models). In this paper, we focus on the latter. The largest uncertainty comes from stellar models. Despite recent advancements in computational astrophysics, stellar evolution remains a problem that is intractable for multidimensional simulations, due to the extremely long timescales of stellar evolution compared to the dynamical timescales of convective processes. Therefore, we rely on spherically symmetric simulations that incorporate convection using mixing-length theory (MLT) \citep{BohmVitense1958_MLT} models. Other multidimensional effects, such as rotation and binary interactions, are typically ignored, although recently an increasing number of models incorporate these effects in 1D \citep{LC18,Laplace2021_core_binary_stars,Farmer2023_Nucleosynthesis_binary}. The other main source of uncertainty is explosion. Core-collapse supernovae are complex environments where extremely high densities are reached, and neutrino interactions play a crucial role \citep{Mezzacappa2020_nu_transport_review,Janka2025_Review_Long3D,Burrows2021_SN_review,Muller2020_SNReview,Boccioli2024_Review}. Therefore, one must consider uncertainties in the nuclear equation of state at high densities, as well as in neutrino transport processes \citep{Fischer2017_Review_EOS_nu}. Finally, multidimensional effects have been shown to be the key ingredient for shock revival \citep{Couch2015_turbulence,Abdikamalov2015_turb_SASI_3D,Radice2016,Radice2018_turbulence} and, therefore, for the successful explosion of a star.

Most predictions for CCSN yields currently rely on relatively simple explosion models to avoid the complexities of CCSN simulations. Moreover, stellar evolution uncertainties are rarely addressed. The goal of this paper is to provide nucleosynthesis yields calculated using a sophisticated explosion model, with state-of-the-art neutrino transport and nuclear equation-of-state (EOS) physics. Moreover, this is one of the few studies where a self-consistent supernova model is used to explode progenitors calculated using different stellar evolution codes. For this purpose, we describe the pre-supernova progenitors, the supernova explosion model, and the explosive nucleosynthesis calculations in Sect.~\ref{sec:methods}. We present various explosion properties in Sect.~\ref{sec:expl_prop}. We discuss the nucleosynthesis, with particular focus on radioactive isotopes in Sect.~\ref{sec:results_NSE_Ye_radio}. We discuss explosion energies and remnant masses in Sect.~\ref{sec:obs}. We present the yields and compare our results to previous work in Sect.~\ref{sec:expl_prop}. Finally, we summarize our discussion in Sect.~\ref{sec:conclusions}.

\section{Methods} 
\label{sec:methods}

All explosive nucleosynthesis calculations rely on two crucial ingredients: (1) the pre-supernova and (2) the explosion models. In this work, we explored the uncertainties related to both ingredients. To address the uncertainties related to the pre-supernova model, we adopted sets of progenitors calculated using three of the most popular stellar evolution codes available:  \verb|KEPLER| \citep[][hereafter WH07]{WH07}, \verb|FRANEC| \citep[][hereafter LC18]{LC18}, and \verb|MESA| \citep[][hereafter F23]{Farmer2023_Nucleosynthesis_binary}. To address the uncertainties related to the explosion model, we compared our results, obtained with a neutrino-driven explosion, with previous results. In these studies, the explosion of the same exact progenitors was manually induced by means of piston, thermal, or kinetic bomb prescriptions and, in one case, by artificially increasing neutrino heating in a different neutrino-driven explosion model.

\subsection{The KEPLER models (WH07)} 
\label{subsec:wh07}
The WH07 models are 32 single-star progenitors at solar metallicity \citep[$Z = 0.0149$,][]{Lodders2003_SolAbu}, with zero-age main-sequence (ZAMS) masses from $12~M_\odot$ to $33~M_\odot$ in steps of one solar mass, plus stars of 35, 40, 45, 50, 55, 60, 70, 80, 100, and 120 $M_\odot$. They were evolved using a coupled 19-isotope network during the main-sequence and post-main-sequence phases, which was replaced by a 128-isotope network in quasi-statistical equilibrium coupled to a small reaction network below magnesium at the end of oxygen burning. The models were then co-processed with an adaptive network up to $\sim 2200$ isotopes. It should be noted that co-processing was only used for nucleosynthesis and did not have any impact on the stellar structure.

\subsection{The FRANEC models (LC18)} 
\label{subsec:lc18}
The LC18 models are 108 single-star progenitors with ZAMS masses of 13, 15, 20, 25, 30, 40, 60, 80, and 120 $M_\odot$, initial rotational velocities of 0, 150, and $\rm 300\ km/s$, and metallicities of $\rm [Fe/H]=0, -1, -2, -3$\footnote{We note here that $\rm [A/B]=log_{10}(X^A/X^B)-log_{10}(X^A_\odot/X^B_\odot)$, where $\rm X^A$ and $\rm X^B$ are the abundances of the species A and B, respectively, and $\rm X^A_\odot$ and $\rm X^B_\odot$ are their solar abundances.}. The adopted solar chemical composition is that provided by \citet{Asplund2009_SolAbu} ($\rm Z=1.345\times10^{-2}$), while in the case of subsolar metallicities an enhancement of $\alpha-$element abundances is included in the scaled solar chemical composition. As a result, the metal fraction corresponding to $\rm [Fe/H]=-N$ is $\rm Z=3.236\times 10^{-(N+2)}$ \citep[see also][]{Roberti2024_z0_expl_nucl}. The nuclear network includes 335 isotopes (from $\rm n$ to \isotope[209]{Bi}), whose temporal evolution is solved simultaneously with the stellar structure and mixing equations.

\subsection{The MESA models (F23)} 
\label{subsec:f23}
The F23 models are 35 single and 31 binary-stripped stars at solar metallicity \citep[$Z = 0.0142$,][]{Grevesse1998_SolAbu}, with ZAMS masses from $11 M_\odot$ to $45 M_\odot$ in steps of one solar mass. Four binary-stripped models with ZAMS masses of 24, 25, 28, and 29 $M_\odot$ did not reach core-collapse due to numerical issues and, therefore, were not included in the study. The binary-stripped stars were all case B systems, where stable Roche-lobe overflow occurs during the Hertzsprung gap. The nuclear network includes 162 isotopes, from $\rm n$ to \isotope[64]{Zn}.

\subsection{Explosion simulations} 
\label{subsec:exp}
The explosion was simulated using the open-source, spherically symmetric code \textsc{GR1D}\footnote{\url{https://github.com/evanoconnor/GR1D}} \citep{OConnor2010} -- with neutrino transport following an M1 moment scheme and neutrino opacities from NuLib \citet{OConnor2015} -- and a time-dependent MLT model for neutrino-driven convection based on \citet{Couch2020_STIR} and \citet{Boccioli2021_STIR_GR}. The spatial grid has 850 radial zones, with a linear spacing of $0.3$~km up to $20$~km and then logarithmically increasing out to the radius at which the density of the progenitor falls below $500\ \rm {g/cm^3}$, which, depending on the progenitor, is typically around $10^{10}\ \rm cm$. The nuclear equation of state adopted is the SFHo \citep{Steiner2013_SFHo}, based on the relativistic mean-field model of \citet{Hempel2010_HS_RMF}. 

GR1D adopts the composition provided by the nuclear equation of state everywhere, i.e., nuclear statistical equilibrium (NSE) abundances of protons, neutrons, alpha particles, and a representative heavy nucleus. At densities below $\sim 10^7$~ g/cm$^3$, it switches to a simple ideal gas in which the ions are assumed to be a pure gas of $\ce{^{56}Ni}$, although the ion contribution to the energy and pressure is negligible at these low densities. The main difference with approaches that solve a nuclear network in regions outside NSE is not so much in the EOS itself but rather in the energy released by nuclear reactions in the pre-shock region, which can aid the explosion \citep{Bruenn2006_conference_burning_shock,Nakamura2014_nucburn_shock,Navo2023_2DSN_ReducedNetwork}. Moreover, post-process nucleosynthesis calculations are also affected, as shown in \cite{Harris2017_postprocess_tracers}. However, that study showed how even a 14-isotope $\alpha$-network can grossly misrepresent nucleosynthesis in regions where the electron fraction differs from $0.5$, compared with a more complete 150-isotope in situ network, which only the ORNL group \citep[][and subsequent works]{Harris2017_postprocess_tracers,Sandoval2021_3D_shk_breakout} uses. With this in mind, we expect the uncertainties in our calculations to be comparable to those adopted in most of the literature.

The neutrino energy grid adopted has 18 energy groups logarithmically spaced from $\sim 1-280$ MeV. Muon and tauon neutrinos and antineutrinos are treated as a single ``heavy-lepton" species, $\nu_x$, since they interact only via neutral-current interactions. The opacities included are emission and absorption of $\nu_e$ and $\bar{\nu}_e$ on nucleons and a representative heavy nucleus, pair production, elastic scattering on baryons, inelastic scattering on electrons, and nucleon-nucleon Bremsstrahlung for the production of $\nu_x$, adopting the formalism of \citet{Bruenn1985} and \citet{BRT2006}. Weak magnetism and virial corrections are included following \citet{Horowitz2002} and \citet{Horowitz2017_virial}, respectively. The MLT parameter was fixed at $\alpha_{\rm MLT}=1.51$, based on the calibration performed against 3D simulations by \citet{Boccioli2021_STIR_GR}, \citet{Boccioli2022_EOS_effect}, and \citet{Boccioli2023_explodability}, as well as the most recent bounds on the fraction of failed supernovae \citep{Adams2017_fraction_failed_SN, Neustadt2021_failedSN_frac}.

\subsection{Nucleosynthesis calculations} 
\label{subsec:nucl}
Nucleosynthesis calculations were performed in post-process using the open-source code \textsc{SkyNet} \citep{Lippuner2017_SkyNet_methods}, with a nuclear network of 1500 isotopes up to nuclei with a mass number of 100 (i.e., the same network used for the test-case hydrostatic burning of carbon and oxygen in \citet{Lippuner2017_SkyNet_methods}). The trajectories were extracted from the explosion simulations and continued up to $t_{\rm fin} = 30$ seconds by extrapolating the temperature and density following a power law \citep{Arcones2007_Nucl_nudriven_outflows,Ning2007_r-process_extrap_traj}:
\begin{align}
    T(t) &= T(t_{\rm fin}) (1 + (t - t_{\rm fin}))^{-2/3} \\
    \rho(t) &= \rho(t_{\rm fin}) (1 + (t - t_{\rm fin}))^{-2}.
\end{align}

It should be noted that the extrapolation method can indeed change the yields by up to $10-20 \%$ \citep{Harris2017_postprocess_tracers,Sieverding2023_tracer_backward,Wang2024_Ti44_Fornax} and is therefore an active source of uncertainty. The time post-bounce when the simulation stops varies between $\sim 3 - 7$ seconds, depending on the progenitor structure. In all cases, the peak temperature reached by the shock at the end of the simulation is below $1.5$ GK.

The time resolution of the adopted tracers varies. For the KEPLER models, we maintained a temporal resolution of $1$~ms up to $1$~s post-bounce, and $3$~ms after that. For the other models, we maintained a temporal resolution of $1$~ms up to $500$~ms post-bounce, $5$~ms until $1$~s post-bounce, and $10$~ms after that. As \citet{Harris2017_postprocess_tracers} showed, the temporal resolution is quite important for preserving the thermodynamic history of the tracer particles. We therefore tested whether adopting the same temporal resolution used for the FRANEC and MESA models would change the results of the KEPLER models by at most $1-2 \%$ for Fe-peak isotopes and by orders of magnitude less for other elements.

Neutrino interactions were ignored in this work since negligible differences were observed by \citet{Wang2024_Nucleosynthesis}. However, we expect significant $\nu$-process nucleosynthesis (which is, however, not included in \textsc{SkyNet}) and, to a lesser extent, $\nu$p-process nucleosynthesis to occur \citep{Frohlich2006_nup_process,Pruet2006_winds_nup,Arcones2013_nuwind,Fischer2024_review_nu_nucleosynthesis,Friedland2025_GR_Newt_nup_semian}. We imposed a temperature threshold for the network to switch to NSE of $7$~GK. Notice that the choice of the threshold can indirectly affect the electron fraction $Y_e$ of each trajectory, as discussed in Sect.~\ref{sec:ye}. The innermost tracer was chosen to be the one with the smallest mass coordinate that has an increasing radius at the end of the simulation and that spent at least 20 ms below NSE temperatures.

\section{Results: Explosion properties}
\label{sec:expl_prop}
\subsection{Explosion energies and remnant masses} 
\label{sec:obs}

Explosion energy is one of the most important observables that can constrain the models and, despite being subject to several modeling assumptions that should not be underestimated \citep{Popov1993_AnModel_IIp,Arnett19982_semian_light_curve,Blinnikov1998_SN1993J_radhydro}, can be directly inferred from the light curve. Typically, the simulations presented in this paper were run up to 2.5-seconds after bounce, depending on the model. Therefore, they are very far from shock breakout, when a true explosion energy can be defined. However, it is standard practice in the field to use the so-called diagnostic explosion energy \citep{Buras2006_2D_diag_ene,Marek2009_SASI_diag_ene,Muller2012_2D_GR,Bruenn2016_expl_en,Muller2017_3Dprog_fix_diag_ene}. We used the following definition of the diagnostic energy from \citet{Muller2017_3Dprog_fix_diag_ene}:
\begin{equation}
    E_{\rm diag} = \int_{e_{\rm tot} > 0} \rho e_{\rm tot} dV,
\end{equation}
where $dV$ is the volume element (which includes general relativistic corrections), $\rho$ is the density, and $e_{\rm tot}$ is the total energy per mass. As discussed in \citet{Muller2017_3Dprog_fix_diag_ene}, defining the gravitational energy in general relativity (GR) is not obvious, and we direct the reader to that paper for a more thorough discussion. Essentially, in GR, the most intuitive definition of $e_{\rm tot}$ leads to a double-counting of the gravitational potential. This can be circumvented by subtracting from $e_{\rm tot}$ the Newtonian potential $\Phi_{\rm grav,\ out}$ of the shells outside a given radius r. This leads to
\begin{equation}
    e_{\rm tot} = \alpha \left[ \left(c^2 + \epsilon + P/\rho\right) W^2 - P/\rho \right] -Wc^2 - \Phi_{\rm grav,\ out}(r),
\end{equation}
with
\begin{equation}
    \Phi_{\rm grav,\ out}(r) = \int_r^\infty \frac{G \rho dV}{r'}dr'.
\end{equation}
In the above equations, $\alpha$ is the lapse function, $\epsilon$ is the sum of the internal and turbulent energy of the matter, $P$ is the pressure, $W$ is the Lorentz factor, $c$ is the speed of light, and $G$ is the gravitational constant. The use of the Newtonian potential for the outer shells is justified because the matter contributing to the total energy is located in the exploding region at radii $r \gtrsim 100$~km, where GR corrections are negligible.

The definition of the internal energy is another crucial detail that can significantly change the value of the diagnostic energy. What must enter the expression of $\epsilon$ is the thermal energy, as pointed out by \citet{Bruenn2016_expl_en}. However, the reference point for the internal energy in the equations of state for nuclear matter adopted in supernova simulations typically accounts for the binding energy of nuclei. Therefore, to calculate the thermal energy, this zero point should be properly readjusted. To do that, we shifted the internal energy in our EOS table so that it matches the internal energy of a pure $\ce{^16O}$ gas at a temperature of 0.01~MeV and a density of $10^6$~${\rm g/cm^3}$, calculated using a Helmholtz EOS \citep{Timmes2000_Helm_EOS}.

The explosion energy was then obtained by subtracting from $E_{\rm diag}$ the energy of the overburden \citep{Bruenn2016_expl_en}, defined as the binding energy of the material outside the shock with negative $e_{\rm tot}$:
\begin{equation}
    \label{eq:expl_ene}
    E_{\rm expl} = E_{\rm diag} - E_{\rm ov}.
\end{equation}

\begin{figure}
    \centering
    \includegraphics[width=\columnwidth]{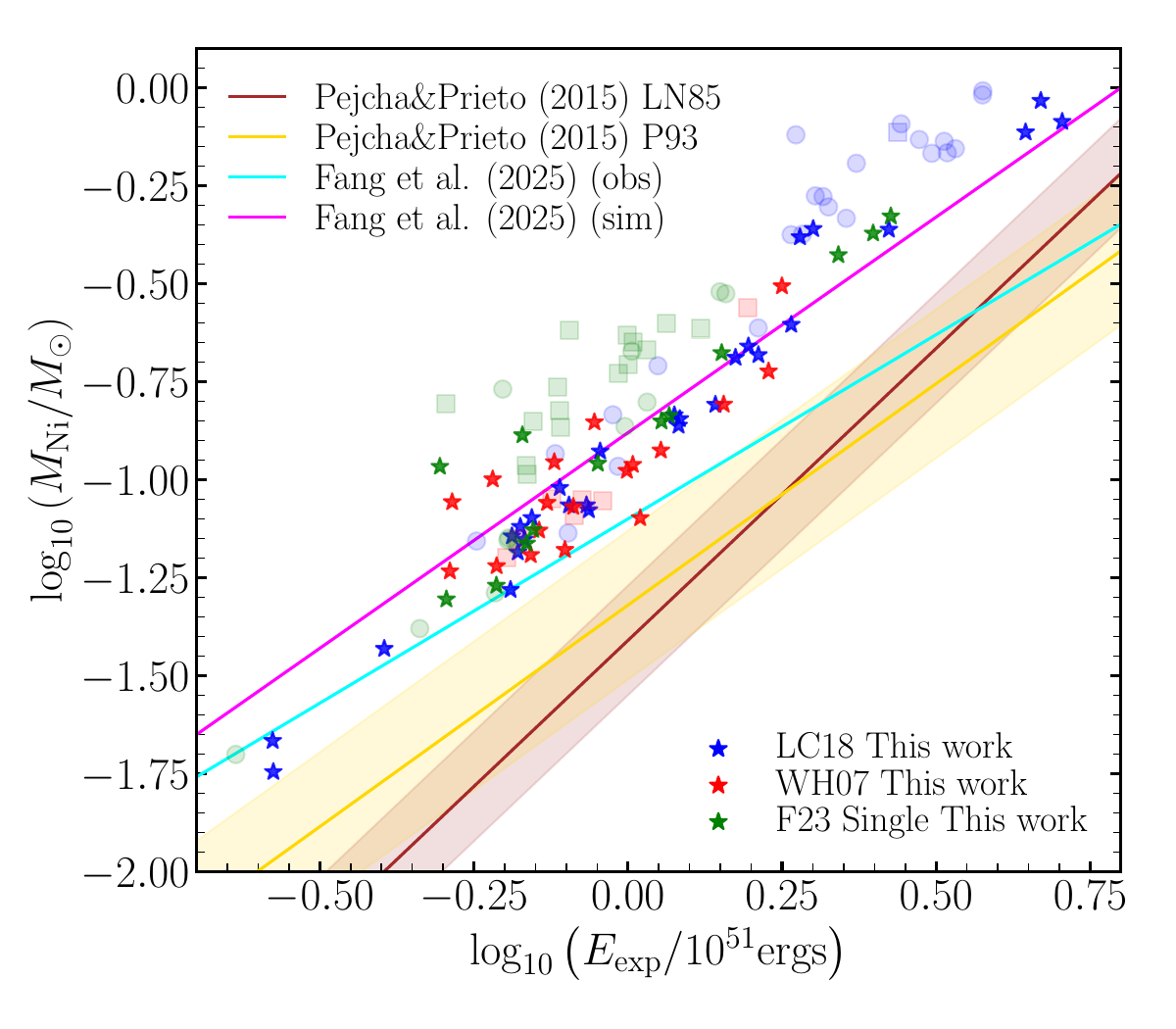}
    \caption{\label{fig:expl_ene_obs} Comparison of explosion energies and ejected $\ce{^56Ni}$ masses obtained from the explosion simulations in this work (indicated by stars) with estimates based on hydrodynamical or semi-analytical modeling of observed light curves. The shaded circles and squares represent simulations of stripped (i.e., with less than $0.01 M_\odot$ of hydrogen in the envelope) and ultra-stripped (i.e., with less than $0.01 M_\odot$ of helium in the envelope) progenitors, respectively. The brown and yellow bands represent the semi-analytical fits based on the light-curve models of \citet{Pejcha2015_typeIIP_fit} using the calibrations of \citet{Litvinova1985_light_curve_params} and \citet{Popov1993_AnModel_IIp}, respectively. The magenta line shows the fit presented by \citet{Fang2025_typeII_light_curve_modeling} to the 2D simulations of \cite{Bruenn2016_expl_en} and to the 3D simulations of \citet{Burrows2024_long3D_kicks_spins}, which is in good agreement with our 1D+ simulations. The cyan line represents a fit derived by \citet{Fang2025_typeII_light_curve_modeling} from observations of 32 type II SNe, in which, instead of assuming a fixed wind model, they used a model grid with different hydrogen envelope masses, thereby reducing the tension between simulations and observations.}
\end{figure}

The explosion energies and ejected $\ce{^56Ni}$ for all the simulated progenitors are shown in Fig.~\ref{fig:expl_ene_obs}. We compare progenitors that have retained a fraction of their hydrogen envelope at the pre-supernova stage to Type II-P SNe. Some outliers in the LC18 set are not shown in Fig.~\ref{fig:expl_ene_obs}. Their main properties are listed in Table~\ref{tab:outliers}. These are progenitors with high compactness $\xi_{2.0} > 0.5$, where the compactness is defined by \citet{OConnor2010} as
\begin{equation}
    \label{eq:compactness}
    \xi_{M} = \dfrac{M/M_\odot}{R(M)/1000\, {\rm km}}.
\end{equation}
Notice that in the original definition, compactness is calculated at bounce, whereas we calculate it using the pre-collapse profile. The two, nevertheless, yield similar values. What is peculiar about these progenitors is that in the simulations, the shock radius is successfully revived, but the protoneutron star (PNS) is so massive that, after losing thermal pressure due to neutrino emission, it leads to black hole formation.

\begin{table}
\caption{List of properties of selected LC18 progenitors that lead to successful explosions and also form a black hole. \label{tab:outliers}}
\begin{tabular}{l|cccccc}
\toprule
Progenitor & $\xi_{2.0}$ & $t_{\rm fin}$ & $M_{\rm PNS}^{\rm fin}$ & $M^{\rm cold}_{\rm NS}$ & $R^{\rm max}_{\rm shock}$ & $E_{\rm expl}$ \\[0.1em]
 &  & (s) & ($M_\odot$) & ($M_\odot$) & (km) & (B) \\
\midrule
060b300 & 0.65 & 0.92 & 2.45 & 2.07 & 7449 & 2.06 \\
080b000 & 0.58 & 0.35 & 2.54 & > 2.08 & 629 & 0.00 \\
080b300 & 0.60 & 0.59 & 2.49 & > 2.08 & 2762 & 0.18 \\
080c000 & 0.58 & 0.64 & 2.49 & > 2.08 & 2586 & 2.22 \\
080d000 & 0.57 & 0.53 & 2.50 & > 2.08 & 2357 & 2.21 \\
\bottomrule
\end{tabular}
\tablefoot{In the first column, the name of the progenitor indicates its ZAMS mass in $M_\odot$ (first three characters), metallicity (a, b, c, and d stands for $z_\odot$, $10^{-1} z_\odot$, $10^{-2} z_\odot$, and $10^{-3} z_\odot$, respectively), and initial rotation in $\rm km/s$ (last three characters). The remaining columns are the compactness (defined in Eq.~\eqref{eq:compactness}, the final time post bounce $t_{\rm fin}$ when the simulation stops (i.e. very close to BH formation), the final baryonic mass of the PNS $M_{\rm PNS}^{\rm fin}$, the gravitational mass of the cold neutron star $M^{\rm cold}_{\rm NS}$corresponding to $M_{\rm PNS}^{\rm fin}$, the maximum shock radius $R^{\rm max}_{\rm shock}$ reached during the simulation, and finally the explosion energy $E_{\rm expl}$ at the end of the simulation defined in Eq.\eqref{eq:expl_ene}.}
\end{table}

As shown in Table \ref{tab:outliers}, the baryonic mass of the PNS corresponds to a cold neutron star with a gravitational mass above the maximum mass allowed by the EOS ($\sim 2.08 M_\odot$)\footnote{Notice that the only exception is the 60 $M_\odot$ progenitor in the first row of Table~\ref{tab:outliers}, which reaches central densities above the upper limit of the table, but we expect it to otherwise be very similar to all the others.}. However, a hot PNS can sustain a mass slightly above the maximum mass of a cold neutron star, and, therefore, for a few hundred milliseconds the added thermal pressure from neutrinos is able to sustain this very large PNS until it cools enough to lead to black hole (BH) formation. 

Based on the discussion above, these progenitors clearly represent edge cases, since the baryonic PNS mass must be just above the maximum allowed mass, but not too far above it; otherwise, the BH forms very quickly and the shock is never revived. However, in a more realistic multidimensional simulation, this could be a much more frequent scenario, due to continued accretion after shock revival. One would therefore still have downflows that contribute to increasing the mass of the PNS. These downflows would be even more pronounced for high-compactness progenitors, since they  develop a larger degree of asymmetry in the explosion \citep{Burrows2024_Phys_correlations}. Therefore, as seen for the progenitors listed in Table~\ref{tab:outliers}, one expects the PNS to collapse to a black hole soon after shock revival, which would halt neutrino emission, produce a weaker explosion, and eject a much smaller amount of $\ce{^56Ni}$. These are progenitors that could give rise to the so-called black hole supernovae (BHSNe), where a successful explosion occurs alongside black hole formation that, depending on how soon after bounce it occurs, can produce very weak to moderately strong explosions of $0.01-2.5\times10^{51}$~erg and leave behind BHs of 20-35 $M_\odot$ \citep{Chan2018_BH_SN_40Msol,Burrows2023_BH_supernova_40Msol,Sykes2024_2D_fallback_SNe,Eggenberger2025_BHSNe_EOS}. In particular, recent simulations of high-compactness progenitors by \citet{Eggenberger2025_BHSNe_EOS} and \citet{Sykes2024_2D_fallback_SNe} found a significant decrease in diagnostic energy right after BH formation, due to halted neutrino emission. In some cases, the diagnostic energy can still slightly increase at later times, but always to values smaller than the initial maximum, in contrast to regular SNe, for which the diagnostic energy increases monotonically until it eventually saturates. Moreover, due to the large fallback of material at late times, most of the $\ce{^56Ni}$ and, more generally, products of explosive Si-burning, are not ejected. Therefore, the estimated explosion energies and $\ce{^56Ni}$ masses obtained for these progenitors from our 1D simulations are likely overestimated. As further confirmation that this could indeed be the case for our outliers, the ZAMS masses of these progenitors are all above 30 solar masses, and their compactness $\xi_{2.0}$ is above 0.5, which are exactly the properties found in the recent simulations of BHSNe mentioned above. 

For all other progenitors, there is a known offset between the explosion models, the estimated explosion energies, and the ejected $\ce{^56Ni}$ masses based on observations of type II-P SNe. This is shown in Fig.~\ref{fig:expl_ene_obs}, where our simulations are compared with the observational fits by \cite{Pejcha2015_typeIIP_fit} and \cite{Fang2025_typeII_light_curve_modeling}. It is important to stress that both explosion energies and ejected masses of $\ce{^56Ni}$ are derived quantities based on semi-analytical and/or radiation-hydrodynamic models used to reproduce the observed bolometric luminosities. Therefore, they are not only be affected by observational uncertainties but also (and, in some cases, mostly) by modeling uncertainties such as those in the progenitor structure, explosion dynamics, and the lack of multidimensional effects. In particular, \cite{Fang2025_typeII_light_curve_modeling} have shown how wind-model assumptions can skew the inference of explosion energies and ejected $\ce{^56Ni}$ masses, and how, by avoiding fits that fix the wind model, one can reduce the tension between multidimensional simulations (magenta line in Fig.~\ref{fig:expl_ene_obs}) and observations (cyan line in Fig.~\ref{fig:expl_ene_obs}). This type of analysis goes beyond the scope of this paper. However, it is important to highlight that the 1D+ models analyzed in this work are in excellent agreement with more sophisticated 2D and 3D models \citep{Burrows2024_Phys_correlations,Bruenn2016_expl_en,Nakamura2015_2D_explodability,Janka2025_Review_Long3D}, as confirmed by the magenta line in Fig.~\ref{fig:expl_ene_obs}.

It is well known that explosion energy directly correlates with the mass of the iron core and, therefore, with the progenitor's compactness \citep{Ebinger2019_PUSH_II_explodability,Janka2017_nuDrivenExpl_Review,Burrows2024_Phys_correlations}. The reason behind this is rather straightforward. More massive iron cores yield higher mass-accretion rates during the stalled-shock phase, since the average densities are higher, which in turn leads to higher neutrino luminosities and energies and, therefore, stronger neutrino heating \citep{Boccioli2025_comp_vs_Qdot}, thereby increasing the explosion energy. Since higher explosion energies are achieved because of higher mass-accretion rates, a natural consequence of this is that higher explosion energies also correlate with the gravitational mass of the cold neutron star left behind by the explosion, which is greater for higher mass-accretion rates and compactness \citep{Boccioli2024_remnant}. To calculate the gravitational mass of the cold neutron star, we first took the baryonic mass of the remnant to be the total mass at densities higher than $10^{12}\ {\rm g/cm^3}$. Then, we converted it to the gravitational mass of the cold neutron star by solving the Tolman--Oppenheimer--Volkoff (TOV) equations. As an alternative definition, we also selected the baryonic mass of the remnant to be the mass coordinate of the innermost tracer and identify differences below 0.1 \%. 

\begin{figure}
    \centering
    \includegraphics[width=\columnwidth]{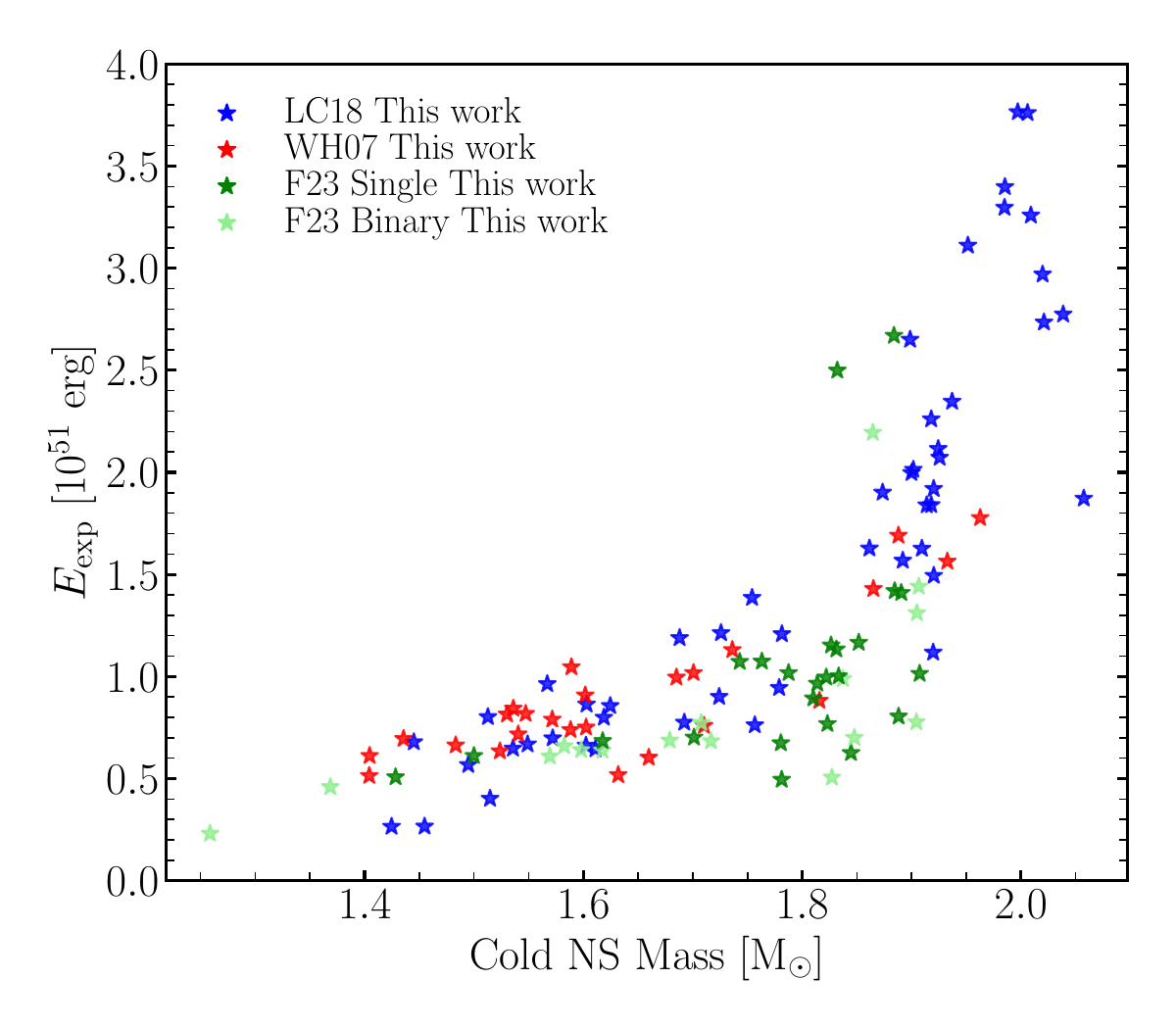}
    \caption{Correlation between the explosion energy and the gravitational mass of a cold neutron star at the end of a simulation, where different colors refer to the different sets of simulations described in Sect.~\ref{sec:methods}. A similar correlation is also found in 3D simulations \citep{Burrows2024_Phys_correlations}. }
    \label{fig:expl_ene_vs_mns}
\end{figure}

To summarize, the mass of the iron core, the compactness, the explosion energy, and the mass of the cold neutron star are all correlated. This is shown in Fig.~\ref{fig:expl_ene_vs_mns}, which can be compared with Fig.~11 of \citet{Burrows2024_Phys_correlations} to highlight once again the excellent agreement between the explosion energies obtained from our 1D+ simulations and those from state-of-the-art 3D simulations.

\subsection{Explodability and explosion dynamics}
\label{sec:expl_dyn}
The advantage of using a full GR simulation with state-of-the-art neutrino transport and a physically motivated model for neutrino-driven convection is that, compared with the legacy methods used to carry out explosive nucleosynthesis calculations (i.e., bomb and piston models), the explosion depends only on the value of the mixing length, which can be calibrated against multidimensional simulations as well as other observables \citep{Boccioli2021_STIR_GR,Boccioli2022_EOS_effect}. Therefore, the explosion arises self-consistently and, as discussed in Sect.~\ref{sec:obs}, so do the mass cut and the explosion energy. Other methods, such as a bomb or a piston, with which the vast majority of explosive yield tables are constructed, require instead that the explosion energy, mass cut, and explosion dynamics are arbitrarily set. Usually, the explosion energy and mass cut are set to reproduce the explosion energy and ejected $\ce{^56Ni}$ mass of SN1987A \citep{Sonneborn1987a}. However, as shown in the previous section, the range of $\ce{^56Ni}$ masses and explosion energies can be quite wide, and therefore a single data point is not enough to represent the full population of supernovae. More recently, more accurate models \cite[see][for a complete list, and the comparison to the present model]{Boccioli2025_comp_vs_Qdot} have been developed, and some have been adopted, with degrees of physical approximations, for nucleosynthesis calculations \citep{Curtis2019_PUSHIII_nucleosynthesis,Sukhbold2016_explodability}.

The differences among 1D explosive nucleosynthesis calculations can therefore be grouped into two main effects: explodability and explosion dynamics. Explodability determines which stars will enrich the ISM. If a star explodes, it will contribute to galactic chemical evolution (GCE) through its winds (i.e., the mass lost during the hydrostatic evolution of the star) and all the mass ejected by the explosion, which is enriched by the products of both explosive nucleosynthesis and pre-supernova nucleosynthesis. If a star does not explode, however, only its winds will pollute the surrounding environment. This is, of course, a simplified picture that does not account for mass ejected in failed explosions \citep{Antoni2023_failed_SNeII}, BHSNe occurrence, or any multidimensional effects. Nonetheless, despite this apparent simplicity, its impact is often entirely neglected by assuming that all stars explode, or it is grossly approximated by setting an arbitrary cutoff of $20-25~M_\odot$ that separates successful explosions (for lower-mass stars) from failed explosions (for higher-mass stars). More recently, however, some efforts toward adopting an approximate neutrino-driven engine to inform explodability and yields have been undertaken \citep{Jost2025_fheat_yields}. With explosion dynamics, we instead define everything that has to do with how the shock wave travels through the star, which depends on the mass cut, the final explosion energy, and the amount of energy injected in the model. In a simplified picture, one can freely vary two of these three quantities, and this is what is typically done in simple bomb and piston models, as we briefly highlight in Sects.~\ref{sec:comparison_WH07},~\ref{sec:comparison_LC18}, and~\ref{sec:comparison_F23}.)

\begin{figure*}
    \centering
    \includegraphics[width=\textwidth]{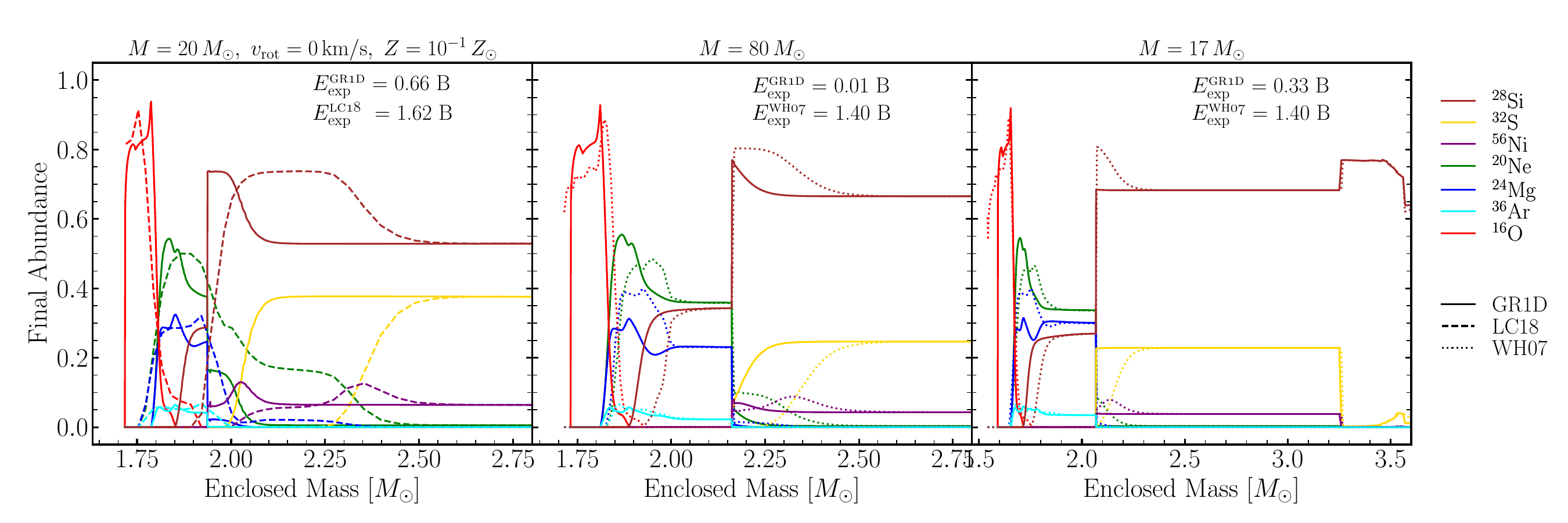}
    \caption{Abundances of selected isotopes after explosive nucleosynthesis for three selected progenitors from LC18 (left) and WH07 (middle and right). The solid lines show the results from the explosion simulated in this work with GR1D, the dashed lines show the bomb model of LC18, and the dotted lines show the piston model of WH07. The progenitor on the left was chosen to be the one for which the mass cut derived with GR1D is closest to that in the original work. The progenitor in the middle was chosen instead to be the one for which the explosion energy derived with GR1D differs most from that in the original work of 1.4 B. The progenitor on the right was chosen to be the one for which the explosion energy derived with GR1D is closest to that in the original work of 1.4 B. As one can see, the left and middle panels are quite similar, showing that the main difference in explosion dynamics is due to the explosion energy, regardless of whether a piston or a bomb is used. When the explosion energies are similar (i.e., the right panel), the differences are smaller.}
    \label{fig:expl_profiles}
\end{figure*}

The explodability obtained in this work is shown in Figs.~\ref{fig:explodability_WH07},~\ref{fig:explodability_LC18},~\ref{fig:explodability_F23} for the WH07, LC18, and F23 sets, respectively. To illustrate how different explodability can be depending on how the explosion is simulated, we added, when available, explodabilities from other 1D studies of the same sets. In Fig. \ref{fig:explodability_WH07}, alongside our explodability, that from \citet{Curtis2019_PUSHIII_nucleosynthesis} (hereafter C19) is also shown. Similarly, in Fig. \ref{fig:explodability_F23}, we also show the explodability derived from the criterion of \citet{Ertl2016_explodability}, which was adopted by \citet{Farmer2023_Nucleosynthesis_binary} as their benchmark. More details on what causes the differences between our explodability and those from C19 and \citet{Ertl2016_explodability} can be found in \citet{Boccioli2025_comp_vs_Qdot}. In general, in contrast with C19 and E16, our model predicts that all high-compactness stars explode, in agreement with the most recent state-of-the-art 3D simulations \citep{Burrows2024_Phys_correlations}, whereas it predicts that lower-compactness stars explode or not depending on how pronounced the density jump at the Si/O interface is \citep{Boccioli2023_explodability,Wang2022_prog_study_ram_pressure}.

Explosion dynamics, by contrast, is less straightforward to compare, given that it is affected by explosion energy, mass cut, and shock velocities, all of which vary from one progenitor set to another and from one explosion simulation to another. The mass cut in WH07 and in F23 was chosen as the point in the star where the specific entropy exceeds 4 (see Sects.~\ref{sec:comparison_WH07} and \ref{sec:comparison_F23} for a more complete discussion). This is essentially a proxy for the Si/O interface, which is mostly in agreement with the mass cut that emerges from our simulations, since the accretion of the Si/O interface through the shock often coincides with shock revival \citep{Boccioli2023_explodability,Boccioli2025_comp_vs_Qdot}. It is important to stress that what matters for the explosion is a drastic change in density, which typically occurs when a pocket of oxygen appears inside the silicon shell. Therefore, one should more correctly refer to a Si-Si/O interface. However, for simplicity and for consistency with previous literature, we refer to this simply as the Si/O interface.

By contrast, LC18 set the mass cut as the point in the star that ensures the ejection of 0.07 $M_\odot$ of $\ce{^56Ni}$ by an explosion that starts inside the iron core \citep[$\sim 1\ \rm M_{\odot}$,][]{Limongi2006_preSN_models}. Their choice was motivated by the observational constraints on the ejected $\ce{^56Ni}$ mass in SN1987A. However, as mentioned above in the context of explosion energies, the range of $\ce{^56Ni}$ masses ejected in a supernova explosion is quite wide. Moreover, since LC18 initiated the explosion inside the iron core, a much higher explosion energy of a few Bethe was required to eject all the material above it, which alters the shock velocity and, more importantly for nucleosynthesis, the shock temperature as it travels through the outer mantle. In their case, the explosion was simulated using a bomb, consisting of depositing a certain amount of kinetic energy in the layers just above the mass cut; this is also the explosion method adopted by F23. WH07 instead used a piston model, where an inner boundary is forced to move outward following a specific time-dependent velocity, as outlined in \citet{Woosley1995_EvolutionExplosionMassive}. 

In general, both for the bomb and for the piston cases, the explosion is more energetic than that arising from our simulations, as shown in Fig.~\ref{fig:expl_profiles} for the selected progenitors. More detailed studies have recently compared piston, bomb, and neutrino-driven models \citep{Imasheva2023_bomb_vs_nu,Imasheva2025_piston_vs_bomb_vs_nu} and have generally concluded that both bomb and piston models struggle to reproduce some of the key properties of neutrino-driven explosions, especially the correlations between explosion energy and the mass of the resulting cold neutron star and of the ejected $\ce{^56Ni}$ mass shown in Figs.~\ref{fig:expl_ene_obs} and \ref{fig:expl_ene_vs_mns}. As we show in the remainder of this section, we generally agree with these findings. 

The bomb model of LC18 for the nonrotating $20 M_\odot$ progenitor with a metallicity of
$\rm Z\simeq10^{-1}\ Z_{\odot}$, shown in the left panel of Fig.~\ref{fig:expl_profiles}, yields an explosion energy of $\sim 1.6$~B, compared with the 0.66~B obtained with GR1D. The reason for such a high explosion energy in the LC18 model is that the bomb was placed inside the iron core, whereas in the piston model of WH07 (and, as discussed above, also in the bomb model of F23), the explosion was initiated at the Si/O interface. In our self-consistent simulations with GR1D, the mass cut naturally arises near the Si/O interface. Therefore, in LC18 the shock has to eject one to a few solar masses more than in the other cases, hence yielding higher explosion energies. The mass cut was then placed by hand to force 0.07 $M_\odot$ of $\ce{^56Ni}$ to be ejected, as mentioned above. The higher explosion energy, therefore, will allow the shock to maintain high temperatures over larger and more extended regions of the pre-supernova structure, as clearly shown by the final abundances in the left panel of Fig.~\ref{fig:expl_profiles}. The $\ce{^56Ni}$ production is relatively similar, with LC18 having a peak at slightly lower masses and a more extended tail at higher masses.

A very similar situation occurs for the WH07 progenitor shown in the right panel, showing that indeed, as one would intuitively assume, the higher explosion energy causes much stronger explosions. Notice that in the piston model of WH07 the mass cut is smaller. This is because the $30\ M_\odot$ star has a relatively large compactness and, therefore, in our simulations, it explodes later, after the Si/O interface has been accreted through the shock. This, as explained above, roughly corresponds to the $s = 4$ layer adopted by WH07 as their mass cut. 
It is perhaps even more instructive to analyze the progenitor in the rightmost panel of Fig.~\ref{fig:expl_profiles}, for which the explosion energy in our model is the closest to the piston model of WH07. The final abundances are much closer in the two models, as one expects. However, the piston still gives slightly higher peak temperatures, showing that even when the explosion energy is similar (and in fact our model has a slightly higher explosion energy), the explosion dynamics is somewhat stronger. As a result, the peak temperature decreases less steeply than in our simulations. 

In summary, both bomb and piston models experience two main shortcomings. First, they arbitrarily set the mass cut, explosion energy, and explosion dynamics. This inevitably prevents them from reproducing the robust trends seen in multidimensional simulations and, more importantly, in observations, among explosion energy, $\ce{^56Ni}$ masses, and neutron-star masses, which neutrino-driven explosions obtain. Second, they arbitrarily set the explodability, contrary to neutrino-driven explosions. However, we highlight that explodability remains an actively debated issue, both in 3D and 1D neutrino-driven simulations, as can be clearly seen from the significant differences in Figs.~\ref{fig:explodability_WH07} and \ref{fig:explodability_F23} among this work, C19, and E16. More details on comparisons of 1D simulations can be found in \citet{Boccioli2025_comp_vs_Qdot}, who have also shown that the 1D+ model adopted in this work is in excellent agreement with both F{\sc{ornax}} \citep{Vartanyan2023_100_2D,Burrows2024_Phys_correlations} and \verb|FLASH| \citep{Eggenberger2021_EOS_dependence_GW_2D,Eggenberger2025_BHSNe_EOS,Li2024_2D_EOS_GW_thesis, Andresen2024_GreyM1} multidimensional simulations, whereas all other 1D models lack validation by 3D simulations.

\begin{figure}
    \centering
    \includegraphics[width=\columnwidth]{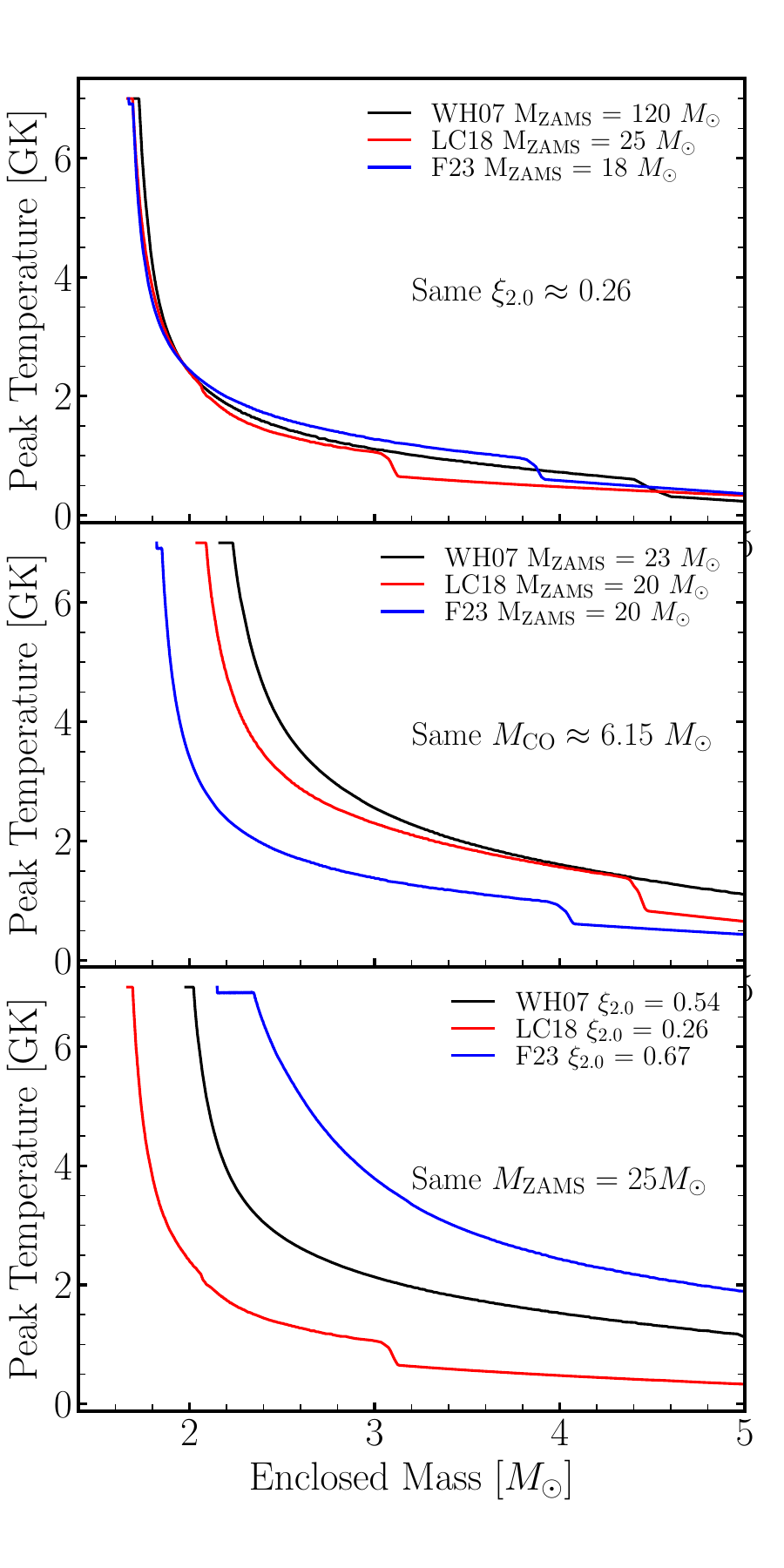}
    \caption{Top: Maximum temperature reached by each mass shell for three different progenitors: a $120 M_\odot$, a $25 M_\odot$, and a $18 M_\odot$ model from WH07 (black), LC18 (red), and F23 (blue), respectively. All are single, nonrotating, at solar metallicity, and all have $\xi_{2.0} \approx 0.26$. Middle: Three progenitors with similar $M_{\rm CO} \approx 6.15$. The WH07 progenitor is the $23 M_\odot$ model, the LC18 progenitor is the $\rm 020b300$ model (following the naming convention in Table~\ref{tab:outliers}), and the F23 model is the $20 M_\odot$ single-star progenitor. Bottom: Three single-star, nonrotating, solar metallicity $25 M_\odot$ models. }
    \label{fig:comparison_tpeak}
\end{figure}

\subsection{Impact of stellar models on the explosion}
One of the largest sources of uncertainty in characterizing the explosion and nucleosynthetic footprint of a massive star is the stellar evolution itself. This paper focuses on the impact of post-collapse physics (e.g., explodability and explosion dynamics, as discussed in the previous section) on a given set of progenitors. However, it is important to point out that the largest source of uncertainty remains stellar evolution, i.e., how each given set of progenitors was constructed. Since we adopted three different models (WH07, LC18, and F23), we briefly show how large the differences in explosive nucleosynthesis are among the three sets. To this end, we analyze the $25 M_\odot$ single-star, nonrotating model at solar metallicity for all three sets, since it is the only one that explodes in all cases. 

The peak temperature as a function of enclosed mass for the $25 M_\odot$ model is shown in the bottom panel of Fig.~\ref{fig:comparison_tpeak} for all three sets. Despite all models being nominally the same progenitor, the setups and assumptions adopted in each stellar model calculation differ, resulting in different stellar structures. These discrepancies may arise from the choice of isotopes and reaction rates included in the nuclear network, or from the implementation of NSE or quasi-statistical equilibrium (QSE) approximations, which become particularly relevant during the final evolutionary stages of massive stars \citep{Roberti2025_SPAr}. Moreover, different treatments of convection can also have a significant impact on the density and compositional structure of the progenitor star. This is evident by examining the compactness $\xi_{2.0}$ of each star, which is quite different, with a value of $0.54, 0.26, 0.67$ for the WH07, LC18, and F23 progenitor, respectively. Since compactness determines the mass cut \citep{Boccioli2024_remnant}, it is not surprising that the peak temperatures are different. Moreover, the shape of the curve is also different. The same holds if, instead of stars with similar ZAMS mass, one compares stars with similar carbon-oxygen core mass $M_{\rm CO}$, as shown in the middle panel of Fig.~\ref{fig:comparison_tpeak}. If one, however, compares stars with similar compactness instead of similar ZAMS mass, as shown in the top panel of Fig.~\ref{fig:comparison_tpeak}, then the peak temperatures are much more similar because the stellar structure, explosion energy, and mass cut are also similar. 

The differences in stellar structure among the stars with the same ZAMS mass also reflect differences in the intrinsic chemical composition of each shell and in the sizes of the shells themselves. Therefore, the pre-supernova and, as a consequence, the explosive nucleosynthesis are quite different. We can thus conclude that the assumptions adopted by different stellar evolution codes dominate the uncertainty in the nucleosynthetic signatures of the stellar models. Even for stars with similar compactness and similar peak temperatures, the burning histories differ and therefore modify the chemical composition of the models.

\begin{figure}
    \centering
    \includegraphics[width=\columnwidth]{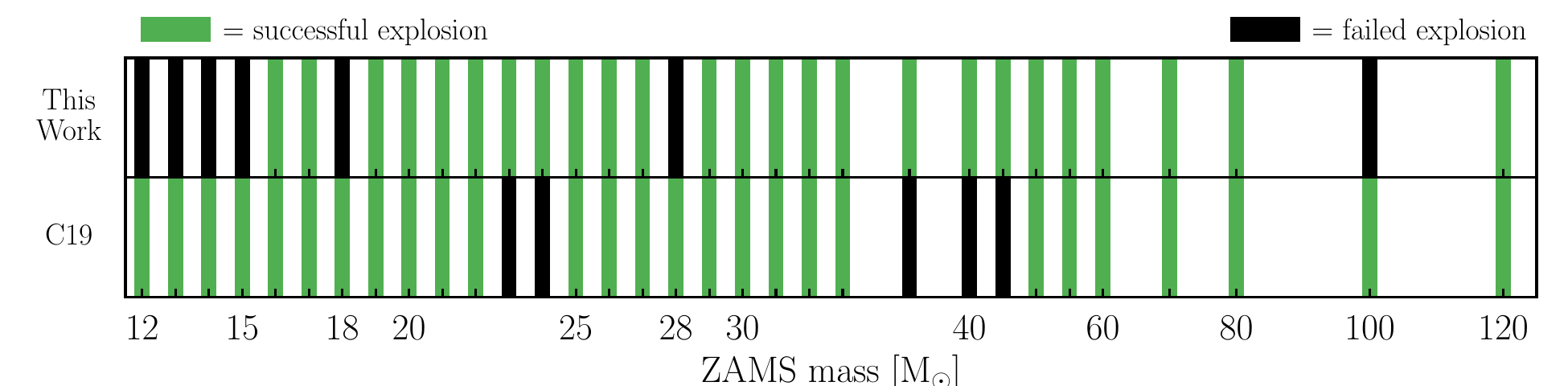}
    \caption{Explodability of progenitor stars from \citet{WH07}, as obtained in this work (top) and by \citet{Curtis2019_PUSHIII_nucleosynthesis} (bottom). Successful explosions, defined as simulations in which the shock is successfully revived and crosses 500 km, are shown as green bands. Failed explosions are shown as black bands.}
    \label{fig:explodability_WH07}
\end{figure}

\begin{figure}
    \centering
    \includegraphics[width=\columnwidth]{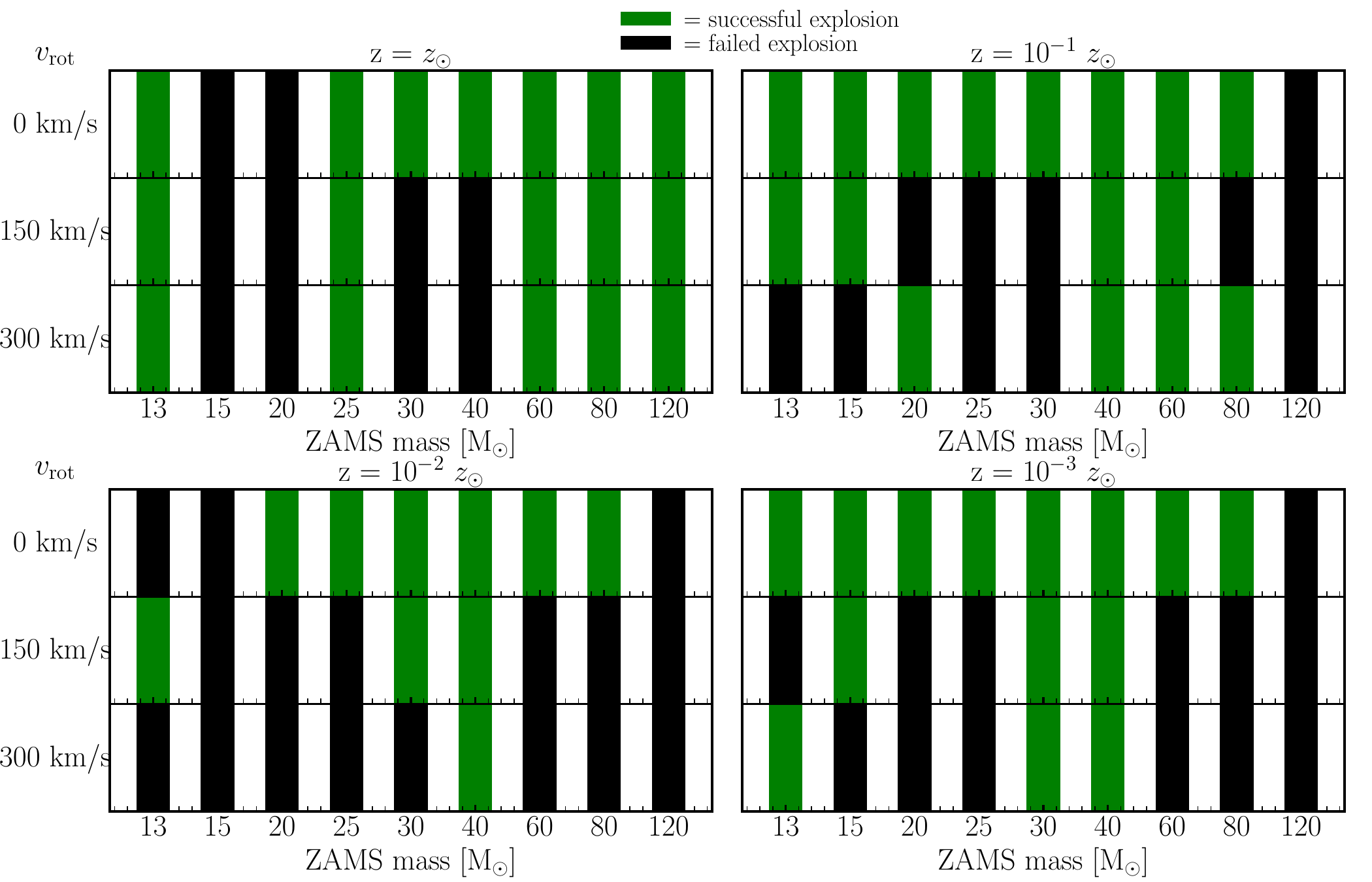}
    \caption{Explodability of progenitor stars from \citet{LC18}, as obtained in this work, shown as a function of initial rotational velocity at the beginning of the main sequence and of metallicity. Successful explosions, defined as simulations in which the shock is successfully revived and crosses 500 km, are shown as green bands. Failed explosions are shown as black bands.}
    \label{fig:explodability_LC18}
\end{figure}

\begin{figure}
    \centering
    \includegraphics[width=\columnwidth]{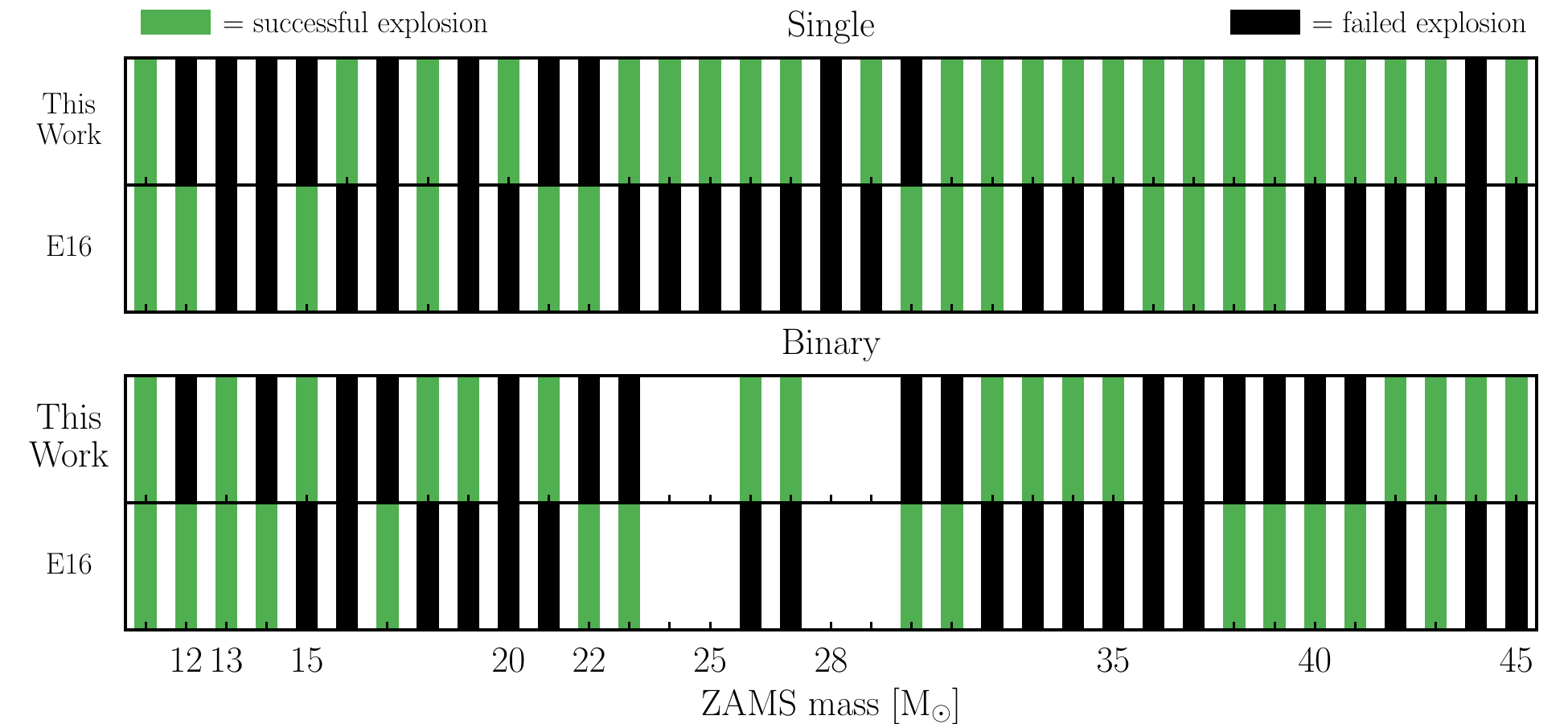}
    \caption{Explodability of single (top) and binary-stripped (bottom) progenitor stars from F23, as obtained in this work and as obtained by F23 using the explosion criterion from \citet{Ertl2016_explodability}, indicated by the label ``E16." Successful explosions, defined as simulations in which the shock is successfully revived and crosses 500 km, are shown as green bands. Failed explosions are shown as black bands. Notice that the binary-stripped progenitors of 24, 25, 28, and 29 $M_\odot$ are missing because, as mentioned in F23, they failed to converge and did not reach the collapse of the iron core.}
    \label{fig:explodability_F23}
\end{figure}

\section{Results: NSE transition, $Y_e$, and radioactive nuclei}
\label{sec:results_NSE_Ye_radio}

\subsection{The role of $Y_e$} 
\label{sec:ye}
An important quantity that can drastically change nucleosynthesis in CCSNe is the electron fraction $Y_e$. Due to neutrino interactions occurring close to the PNS, $Y_e$ could, in principle, be quite different from 0.5. Values of $Y_e < 0.5$ would favor weak $r-$process nucleosynthesis \citep{Meyer1992_rprocess_SN,Woosley1994_rprocess_SN,Wanajo2013_rprocess_PNS}, allowing for elements up to the first or, in rare cases, second $r-$process peak, to be produced. Values of $Y_e > 0.5$ would instead favor the production of neutron-deficient nuclei through, for example, the $rp$-process \citep{Schatz1998_rp_process, Wanajo2006_rp_proc} or the $\nu p$-process \citep{Frohlich2006_nup_process,Pruet2006_winds_nup}. In recent multidimensional simulations \citep{Wang2024_Nucleosynthesis, Wang2023_nuwinds}, it has been shown that a wide range of $Y_e$ can be achieved, due to neutrino interaction and, crucially, asymmetric explosions, which enable more matter to be exposed to neutrino fluxes near the PNS and, for low-mass progenitors, early explosions that prevent neutrinos from increasing $Y_e$ in the early $\nu$-driven wind, thereby enabling a significant weak r-process.

\begin{figure}
    \centering
    \includegraphics[width=\linewidth]{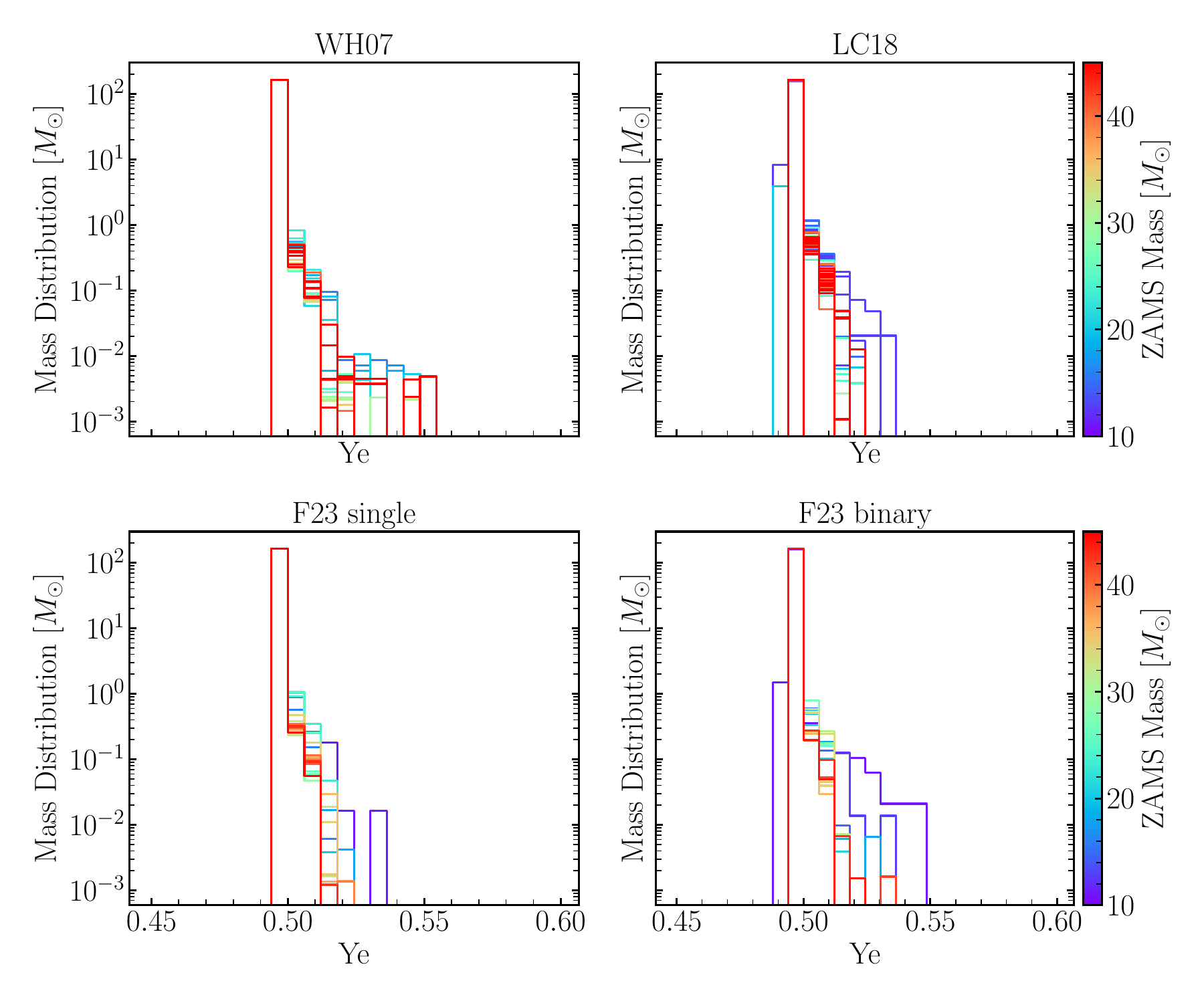}
    \caption{$Y_e$ distributions of the ejecta color-coded by the ZAMS mass of the progenitor star for each of the three sets of simulations. Notice that we separated the single and binary stars of F23 in the bottom two panels. Lower-mass stars reach higher $Y_e$, as also seen in 3D simulations \citep{Melson2015_9.6_expl,Muller2019_3D_lowmass_binary,Stockinger2020_3D_low_mass_to_breakout,Sandoval2021_3D_shk_breakout,Wang2024_LowMass_3D} }
    \label{fig:Ye_distr}
\end{figure}

Because they lack these key multidimensional effects, spherically symmetric simulations cannot reproduce these same conditions. In our models, we do not find any significant weak $r-$process, since $Y_e$ never drops significantly below 0.48-0.49. However, since we consistently evolve neutrino interactions (contrary to the studies of WH07, LC18, F23, and E16), our inner mass shells can achieve $Y_e$ of up to 0.55 in some cases, especially for low-mass models, which is consistent with other 3D studies \citep{Melson2015_9.6_expl,Muller2019_3D_lowmass_binary,Stockinger2020_3D_low_mass_to_breakout,Sandoval2021_3D_shk_breakout,Wang2024_LowMass_3D}. Therefore, the same (or, at least, very similar) nucleosynthesis observed in proton-rich ejecta in 3D is also observed in our models. The only difference is what is usually referred to as reheating due to secondary shocks at late times \citep{Sieverding2023_3D_nucleosynthesis}, which can sometimes reheat matter from $\sim 1.5$~GK to temperatures above $\sim 2.5$~GK, thereby causing $\alpha$-chain reactions that produce $\ce{^44Ti}$. The key difference still remains the amount of proton-rich ejecta, which in 1D can be a factor of 5-10 lower compared to 3D. The $Y_e$ distributions of all the simulated models are shown in Fig.~\ref{fig:Ye_distr}.

Another key aspect of explosive nucleosynthesis is setting the temperature threshold above which matter is considered to be in NSE. In the literature, typical values have been chosen to be anywhere between 5 and 10~GK, mostly because tabulated nuclear reaction rates are not properly defined above 10 GK. Therefore, solving the simplified NSE equations introduces a much smaller uncertainty than attempting an extrapolation of the reaction rates. For stellar nucleosynthesis calculations, this is only relevant in the last Si-burning stages, and one typically would expect that higher temperature thresholds would be more realistic, since a sufficiently large network should be able to naturally reproduce the NSE results even at very high temperatures. For explosive nucleosynthesis, however, this is not true, although it is often assumed to be. The key difference is the role of neutrinos. If temperature thresholds are set too high, nucleosynthesis calculations for tracers never exposed to high temperatures would start from the pre-collapse composition. For example, choosing 10 GK as the NSE threshold means that tracers not heated to this temperature undergo post-processing from the initial composition, i.e., $Y_e \lesssim 0.5$. This ignores all prior neutrino interactions, significantly altering their $Y_e$. Even when included in network calculations, as in many nucleosynthesis codes \citep{Hix1999_XNet_methods,Frohlich2006_CFNET_methods,Lippuner2017_SkyNet_methods,Reichert2023_WinNet_methods_paper}, neutrino interactions remain far less sophisticated than explicitly solving the Boltzmann equation in neutrino transport, often yielding different values of $Y_e$ than simulations. A more detailed nucleosynthesis analysis would be necessary to determine how the $Y_e$ in the network and the simulation differ as a function of the NSE threshold. This is, however, beyond the scope of this paper and is left for future work.

In practice, this leads to moderate-to-severe differences in selected isotopes, particularly those produced during complete silicon burning ($\rm T\geq 5~GK)$ near the PNS. A clear example of this is $\ce{^44Ti}$. To test how sensitive our results are to the chosen NSE threshold, we also ran nucleosynthesis calculations with an NSE threshold set at 10~GK for the WH07 progenitors. Unfortunately, SkyNet cannot handle NSE thresholds higher than 7-8 GK \citep{Lippuner2017_SkyNet_methods}. Therefore, as proof of principle, we took the trajectories with peak temperatures between 7 to 10 GK, set a ceiling of 7 GK, and initialized the abundances to pre-supernova values. Although not equivalent to a proper nucleosynthesis calculation between 7 and 10 GK, this approach still highlights the impact of ignoring the $Y_e$ evolution for these tracers. More accurate methods should be adopted for a thorough study. This is nevertheless beyond the scope of this paper. 

\begin{figure}
    \centering
    \includegraphics[width=\linewidth]{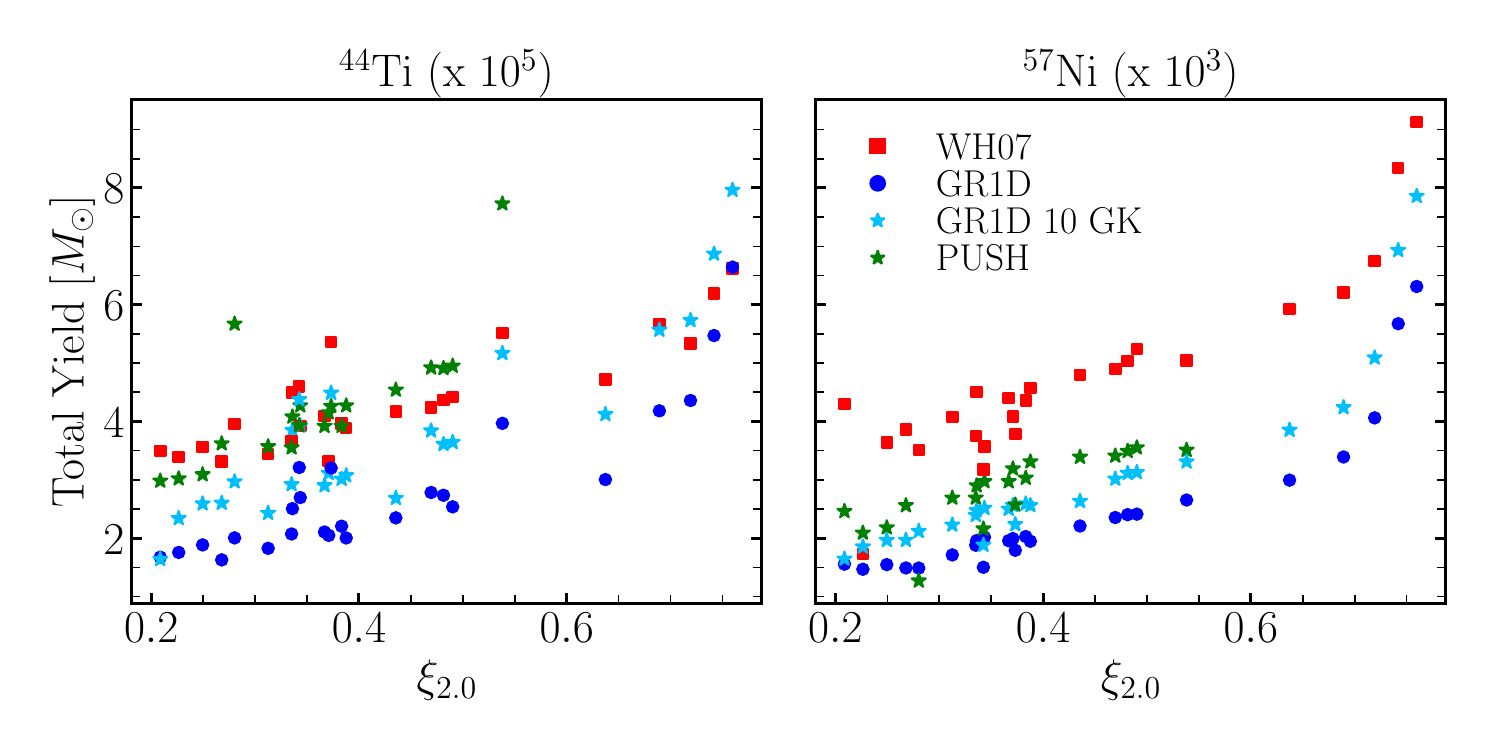}
    \caption{Total yields of $\ce{^44Ti}$ (left) and $\ce{^57Ni}$ (right) multiplied by factors of $10^5$ and $10^3$, respectively. The cyan symbols refer to a nucleosynthesis calculation that mimics one in which 10 GK is used as the NSE threshold (although see the text for a detailed explanation of how that was actually done). The increase in the yield of both isotopes compared with the standard run (blue symbols) shows that part of the discrepancy with the C19 results lies in the choice of the threshold. }
    \label{fig:Ti44_10Gk}
\end{figure}

We compare our standard nucleosynthesis results with this approximate 10 GK threshold in Fig.~\ref{fig:Ti44_10Gk} alongside the results from WH07 and C19. In the latter study, an NSE threshold of $\sim 10$~GK was used; however, unlike our work, it included neutrino emission and absorption during post-processing. As clearly shown, our $\ce{^44Ti}$ and $\ce{^57Ni}$ yields using the 10 GK approximate threshold are much closer to the C19 results than our standard calculation. This is because with a higher NSE threshold, more inner tracers are initialized at the pre-supernova $Y_e = 0.5$. Therefore, this leads to higher abundances of several $\alpha$-elements such as $\ce{^44Ti}$.  

As shown by \cite{Harris2017_postprocess_tracers} for a 21-isotope network, modifying the NSE threshold has a further consequence: an overly low transition temperature prevents the capture of $\alpha$ particles produced during the NSE, leading to $\alpha$-rich freeze-out and reduced $\alpha$-element abundances.
Therefore, if one verifies that the $Y_e$ from post-processing reproduces the $Y_e$ in the simulation, a higher NSE threshold should not modify the resulting abundances. However, given the far more sophisticated neutrino interactions adopted in the simulations, post-processing -- even when neutrino interactions are present -- does not guarantee identical $Y_e$ values. Finally, it is important to highlight that WH07 yields are even larger, as a direct consequence of their higher explosion energies, as explained in previous sections.

\subsection{Radioactive nuclei} 
\label{sec:radioactives}
In this section, we briefly discuss the two most important radioactive nuclei ejected during the explosion: $\ce{^56Ni}$ and $\ce{^44Ti}$. The former decays to $\ce{^56Co}$ with a half-life of $\sim 6$~days, which subsequently decays to $\ce{^56Fe}$ with a half-life of $\sim 77$~days. The latter decays to $\ce{^44Sc}$ with a half-life of $\sim 60$~years, which subsequently decays to $\ce{^44Ca}$ with a half-life of $\sim 4$~hours. This means that the former is responsible for powering the early light curve of CCSNe, whereas the latter is an important element for observations of supernova remnants on timescales of tens to hundreds of years.

\begin{figure}
    \centering
    \includegraphics[width=\linewidth]{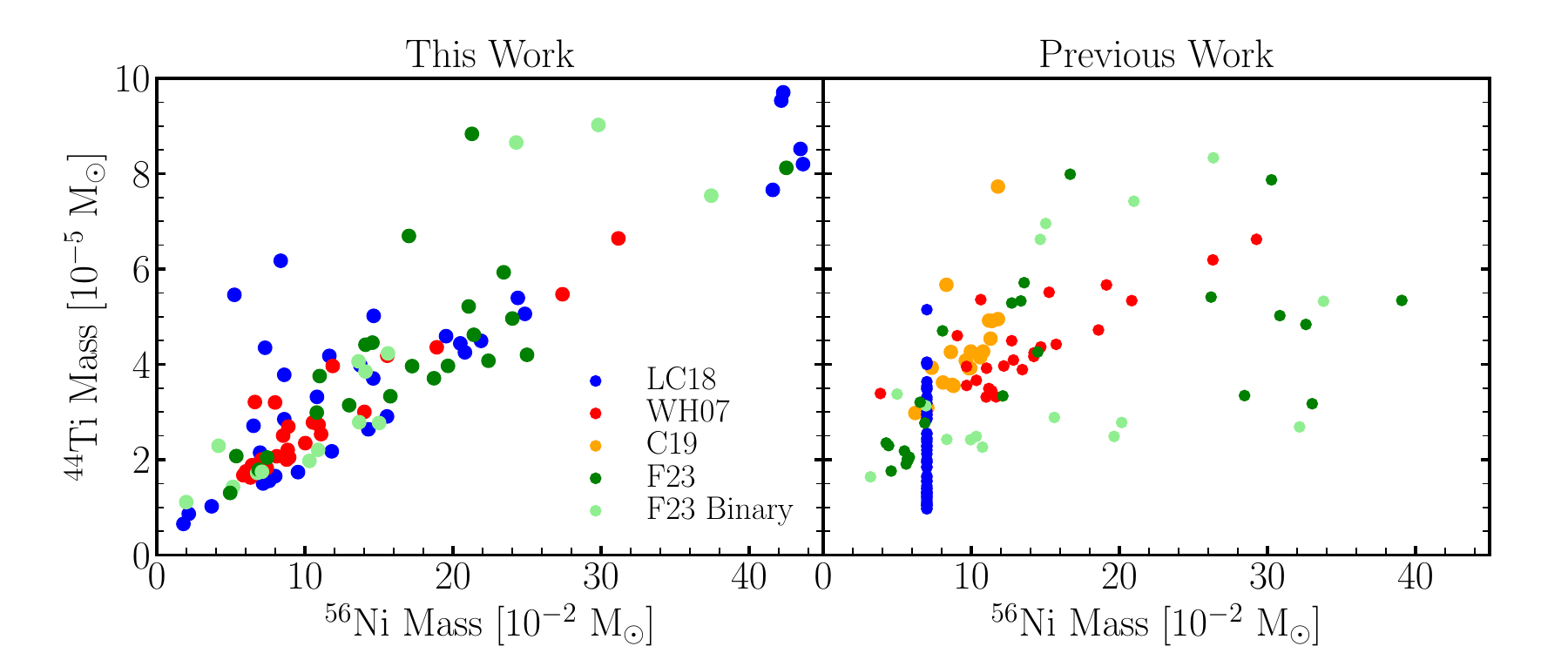}
    \caption{Left: Total yields of $\ce{^44Ti}$ and $\ce{^56Ni}$ for the nucleosynthesis calculations carried out in this work. The different colors denote the different sets adopted in this paper. Right: Same quantities, but with the nucleosynthesis calculations performed as in the original LC18, F23, and WH07 papers. In addition, we also show the results from C19, who exploded the WH07 progenitors. Notice that, in their bomb explosion model, LC18 fix the amount of $\ce{^56Ni}$ ejected to 0.07 $M_\odot$. }
    \label{fig:Ti44_vs_Ni56}
\end{figure}

As mentioned in the previous sections, the nucleosynthesis of $\ce{^44Ti}$ during incomplete Si-burning is quite sensitive to the peak temperature reached by each tracer. Therefore, secondary shock in multidimensional simulations that reheat the matter after initial expansion boosts its production by a factor of a few \citep{Sieverding2023_3D_nucleosynthesis,Wang2024_Nucleosynthesis}. Nonetheless, it is still insightful to compare the yields obtained in this work with those from previous 1D studies. This comparison is shown in Fig.~\ref{fig:Ti44_vs_Ni56}. Focusing on the left panel first, which shows the yields obtained in this work, we can conclude that there is no significant difference in the trend among the three different sets. Moreover, there is no obvious dependence on metallicity or initial rotation in the LC18 progenitors, or between single and binary stars in the F23 progenitors. This is not surprising, since only the final structure (i.e., compactness) of the star is responsible for determining the amount of mass that undergoes complete and incomplete silicon burning, as shown in Fig.~\ref{fig:Ti44_10Gk}, which in 1D is the only production site of $\ce{^56Ni}$ and $\ce{^44Ti}$. 

However, the right panel exhibits much more pronounced scatter among the points. The most strikingly different result is the one from LC18, which is simply the consequence of imposing a fixed amount (i.e., $0.07 M_\odot$) of $\ce{^56Ni}$ to be ejected in all models. Both C19 and WH07 report higher $\ce{^44Ti}$ yields than our findings. For C19, the reason for this was discussed in Sect.~\ref{sec:ye} and most likely stems from the higher NSE threshold. For WH07, the reason is quite similar: they do not include neutrinos at all, leaving all tracers at $Y_e$ $\sim 0.5$, which, as illustrated in the previous section, increases $\ce{^44Ti}$ abundances. For F23, the situation is more complex: their $\ce{^44Ti}$ yields show a very wide spread, though the reason remains unclear since only the final yields for those models are publicly available. 

It is also worth pointing out the six models at the top left of the left panel, which have enhanced $\ce{^44Ti}$ than other models with comparable amounts of $\ce{^56Ni}$. Models exhibiting this exact same behavior also appear in \citet{Sukhbold2016_explodability} and \citet{Curtis2019_PUSHIII_nucleosynthesis} (see top left of Fig. 3 in \citet{Sieverding2023_3D_nucleosynthesis}). These are progenitors that experienced the so-called carbon-oxygen (C--O) shell mergers \citep{Andrassy2020_3D_CO_merger,Rizzuti2023_3D_CO_merger,Roberti2025_SPAr}, i.e., complete (or sometimes incomplete; \hbox{Roberti \& Boccioli, in prep.)} mixing of the carbon and oxygen shells a few hours to a few days before collapse. During the shell merger, typical shell oxygen burning products are synthesized alongside significant amounts of $\ce{^44Ti}$, odd-Z elements (P, Cl, K, and Sc) and p-nuclei \citep{Chieffi2017_Ti44,Roberti2023_GammaProcess1,Roberti2025_CO_merger_K_Sc,Boccioli2026_ProdHeavAele44TiCasCompAbun}. Notice that, if a C--O merger occurs, the material outside the Si-shell quickly expands, decreasing its density. Therefore, when the shock reaches these layers, it lacks the energy to heat the material to high-enough temperatures to modify its composition \citep{Roberti2024_Gammaprocess2}. The net effect is that the pre-supernova abundances in the carbon and oxygen shells are largely ejected without being reprocessed \hbox{(Roberti \& Boccioli, in prep.) in these progenitors}. 

\section{Results: Yields and comparison with previous work}
\label{sec:comparison}
In this section, we extensively compare our work with previous studies that have adopted the same progenitors to perform explosive nucleosynthesis calculations. The goal is to highlight how different explodability and different explosion engines (i.e., a different explosion dynamics) affect the results. Moreover, comprehensive comparisons between nucleosynthesis calculations from sophisticated explosion simulations (e.g., this work) and more simplistic, widely used explosion models remain scarce in the literature. A sample of elemental and radioactive isotope yields is shown in Appendix~\ref{sec:appendix_tables}, and the full tables are available online (see Data Availability section).

\subsection{Comparison with PUSH and WH07}
\label{sec:comparison_WH07}
As mentioned above, the WH07 models analyzed in this work include those from \citet{WH07}, who performed explosive nucleosynthesis calculations using the piston model to simulate shock propagation \citep[see, e.g.,][]{Woosley1995_EvolutionExplosionMassive,Woosley2002_KEPLER_models}. In particular, they provide four different explosion models for each progenitor: (i, ii) setting the mass cut at either the iron core edge or where the specific entropy per baryon is 4; (iii, iv) setting the explosion energy (i.e., kinetic energy of the ejecta at infinity) to 1.4 or 2.4 B.

We compare our results to WH07 models with a 1.4 B explosion energy and mass cut at a specific entropy per baryon of 4, (roughly the Si/O interface) because, as seen in Fig.~\ref{fig:expl_ene_obs}, our explosion energies for all the WH07 progenitors are lower than 2.0 B. We stress again that these widely used nucleosynthesis yields treat explosion energy as a free parameter and is usually fixed for all progenitors, which is, of course, nonphysical as discussed in the previous section. We only compare to the WH07 set where the mass cut is set to the Si/O interface since, as it has extensively been shown \citep{Boccioli2024_FEC+_SiO,Vartanyan2021_Binary_stars_SiO_interface,Boccioli2023_explodability,Lentz2015_3D,Summa2016_prog_dependence_vertex,Vartanyan2018_2D_revival_fittest,Ertl2016_explodability,Wang2022_prog_study_ram_pressure,Tsang2022_ML_explodability}, the accretion of this interface through the shock often corresponds to the onset of the explosion, and therefore its location roughly corresponds to the remnant mass \citep{Boccioli2024_remnant,Liu2021_remnant_LMG,Raithel2018_remnant_mass,Patton2020_popsynth_prescription}.

\begin{figure}
    \centering
    \includegraphics[width=\linewidth]{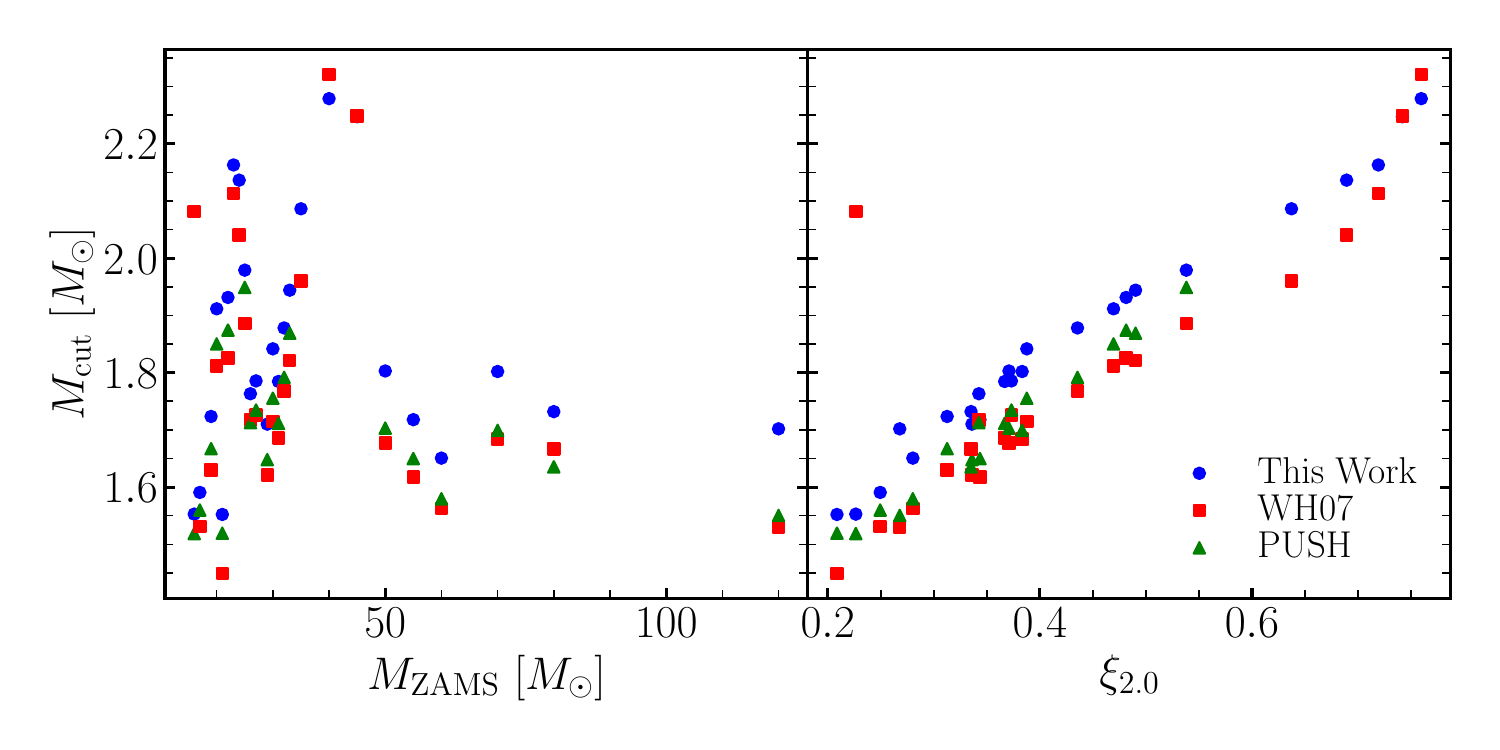}
    \caption{Mass cuts obtained in this work (blue dots) compared with those obtained by WH07 (red squares) and C19 (green triangles) for the explosions of the WH07 progenitors. As expected, the mass cut correlates with compactness (except for one outlier from WH07). Notice that we find mass cuts that are larger than those of C19 and WH07, who instead find comparable values. The reason is that our explosions are triggered some time after the Si/O interface is accreted, whereas in the case of WH07 they force the explosion exactly at the location of the interface. C19 forces the explosion by increasing heavy-neutrino heating and, as a consequence, as soon as the Si/O interface crosses the shock, the explosion occurs much more quickly than in our simulations, and therefore mass accretion stops abruptly. The one outlier from WH07 has a mass cut close to the outer Si/O interface. However, an oxygen pocket is already present much closer to the inner core, hence the large difference with respect to our models. }
    \label{fig:mcut_WH07}
\end{figure}

The location of the mass cut is plotted for all three sets in Fig.~\ref{fig:mcut_WH07} as a function of initial ZAMS mass and pre-supernova compactness (defined in Eq.~\ref{eq:compactness}). The well-known correlation between compactness and mass cut appears in all three sets and is simply the consequence of the dependence of the remnant mass on the Chandrasekhar mass \citep{Timmes1996_NS_BH_birth_mass, Boccioli2024_remnant}. The latter increases with iron core mass and, thus, compactness, since larger iron cores produce more compact stars.

The pre-supernova WH07 models have also been exploded in spherical symmetry using the neutrino-driven PUSH method \citep{Perego2015_PUSH1,Ebinger2019_PUSH_II_explodability,Ebinger2020_PUSH_IV_nucleosynthesis}. In particular, an extensive study of these same progenitors was carried out by \citet{Curtis2019_PUSHIII_nucleosynthesis} (hereafter C19).
The explosion was simulated using a slightly more approximate neutrino transport than that adopted here. Nonetheless, their spectral neutrino transport \citep{Liebendorfer2009_IDSA} accurately follows the evolution of electron neutrinos and antineutrinos. However, the much cruder advanced spectral leakage (ASL) approximation was used for heavy-lepton neutrinos \citep{Perego2016_advanced_leakage}, yielding nonphysically high energies. 

In PUSH, the explosion is triggered via an ``extra-heating" term on the RHS of the energy equation, which depends on the energy and luminosity of heavy-lepton neutrinos and the compactness of the pre-supernova progenitor. See \citet{Perego2015_PUSH1} for more details and Sect. 5.2 of \citet{Boccioli2025_comp_vs_Qdot} for a comparison between \textsc{GR1D+} and PUSH.

The other main difference with the explosion simulations carried out in this work is that the PUSH method also includes a small alpha network in non-NSE regions. We do not expect this to significantly impact results, as \citet{Navo2023_2DSN_ReducedNetwork} showed that in 1D, including a small burning network slightly reduces pre-shock ram pressure and strengthens explosion. Given that the strength of the explosion is, in any case, determined by tweaking the ``extra-heating" in PUSH, we do not expect this to have any impact on the comparison.

For meaningful comparison, we weighted the entire set using a Salpeter initial mass function \citep[IMF,][]{Salpeter1955_IMF} independent of metallicity, rather than comparing each progenitor separately:
\begin{equation}
    \label{eq:IMF}
    P(M_{\rm ZAMS}) \propto M_{\rm ZAMS}^{-\alpha_{\rm IMF}},
\end{equation}
with $\alpha_{\rm IMF} = 2.35$, where $P(M_{\rm ZAMS})$ is the probability of forming a star of ZAMS mass of $M_{\rm ZAMS}$. This demonstrates how explodability influences the overall contribution of the stellar population to ISM enrichment. Yields were thus IMF-weighted, averaged, normalized to 1 (i.e., we calculated the mass fractions of each isotope), and converted to number fractions\footnote{$\rm Y^i=X^i/A^i$, where X is the mass fraction and A is the atomic mass of the nuclear species $i$.} for comparison with WH07 and C19.

As shown in Fig.~\ref{fig:explodability_WH07}, the explodability from our \textsc{GR1D+} simulations differs substantially from that obtained by C19 using PUSH. Notice that WH07 assume everything explodes and therefore provide explosive yields for all progenitors. The main difference in explodability is that, with \textsc{GR1D+}, stars with masses between 12 $M_\odot$ and 15 $M_\odot$ do not explode, whereas with PUSH they do. Moreover, stars around 25 $M_\odot$ and 40 $M_\odot$ do not explode with PUSH, but do with \textsc{GR1D+}. This occurs because PUSH is calibrated such that high-compactness stars -- particularly those around 25 $M_\odot$ and 40 $M_\odot$ -- do not explode (see \citet{Perego2015_PUSH1} for the exact calibration). 

\begin{figure}
    \centering
    \includegraphics[width=\linewidth]{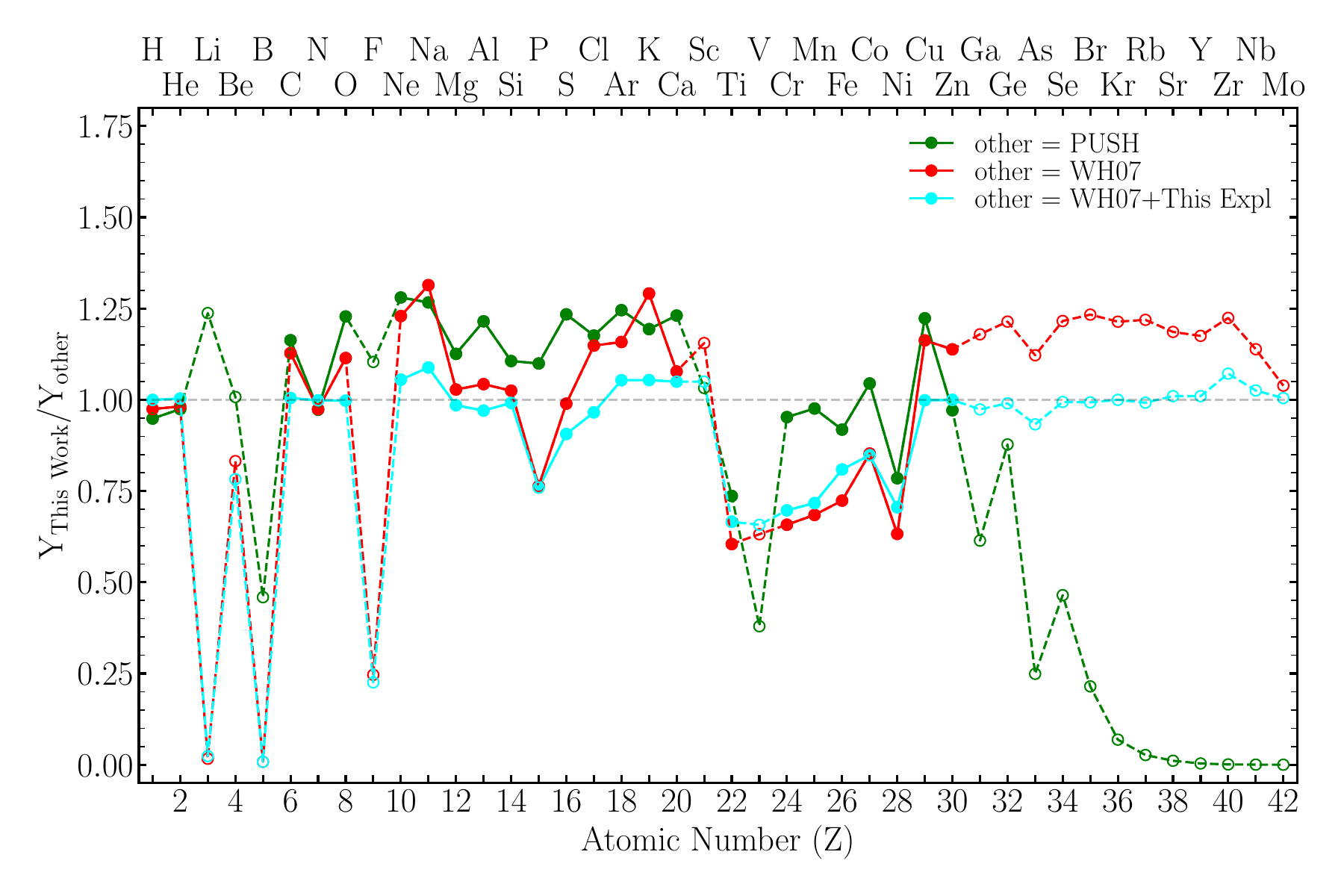}
    \caption{Ratio of the yields obtained in this work to those obtained in previous studies, each weighted by a Salpeter IMF and subsequently normalized to one. The red line shows the ratio of our yields to the original yields from WH07, for which every star explodes. The cyan line is the same as the red, except that we apply our explodability to the yields of WH07. This means that for the 12, 14, 15, 18, and 100 $M_\odot$ stars, which do not explode in our simulations, we only include the winds, whereas for the rest we take the full yields from WH07. The green line shows the ratio of our yields to those by C19. The empty circles and the dashed line connecting them indicate elements for which the total yield is less than $10^{-7} M_\odot$.}
    \label{fig:Element_comparison_WH07}
\end{figure}

To demonstrate the impact of explodability, in addition to comparing our yields to the original ones from WH07, we also compare them to the WH07 yields, but this time imposing \textsc{GR1D+} explodability. Non-exploding progenitors with \textsc{GR1D+} thus contribute only via stellar winds.  Elemental abundance comparisons assuming that everything has decayed back to stability are shown in Fig.~\ref{fig:Element_comparison_WH07}.

\subsubsection{WH07}
Overall, the yields obtained in this work agree with those obtained in WH07 within a factor of $\sim 0.5-1.5$. The first effect we analyze is explodability. If one assumes that everything explodes (i.e., the red line in Fig.~\ref{fig:Element_comparison_WH07}), then our simulations overall produce an increased number of light and trans-Fe elements than WH07 (except phosphorus and fluorine, which deserve separate discussion). By contrast, WH07 produce fewer Fe-peak elements. When we impose our explodability on the WH07 yields (i.e., the cyan line in Fig.~\ref{fig:Element_comparison_WH07}), we find significantly improved agreement for light elements and trans-Fe elements, whereas agreement for Fe-peak elements improves only very slightly.

\begin{figure}
    \centering
    \includegraphics[width=\linewidth]{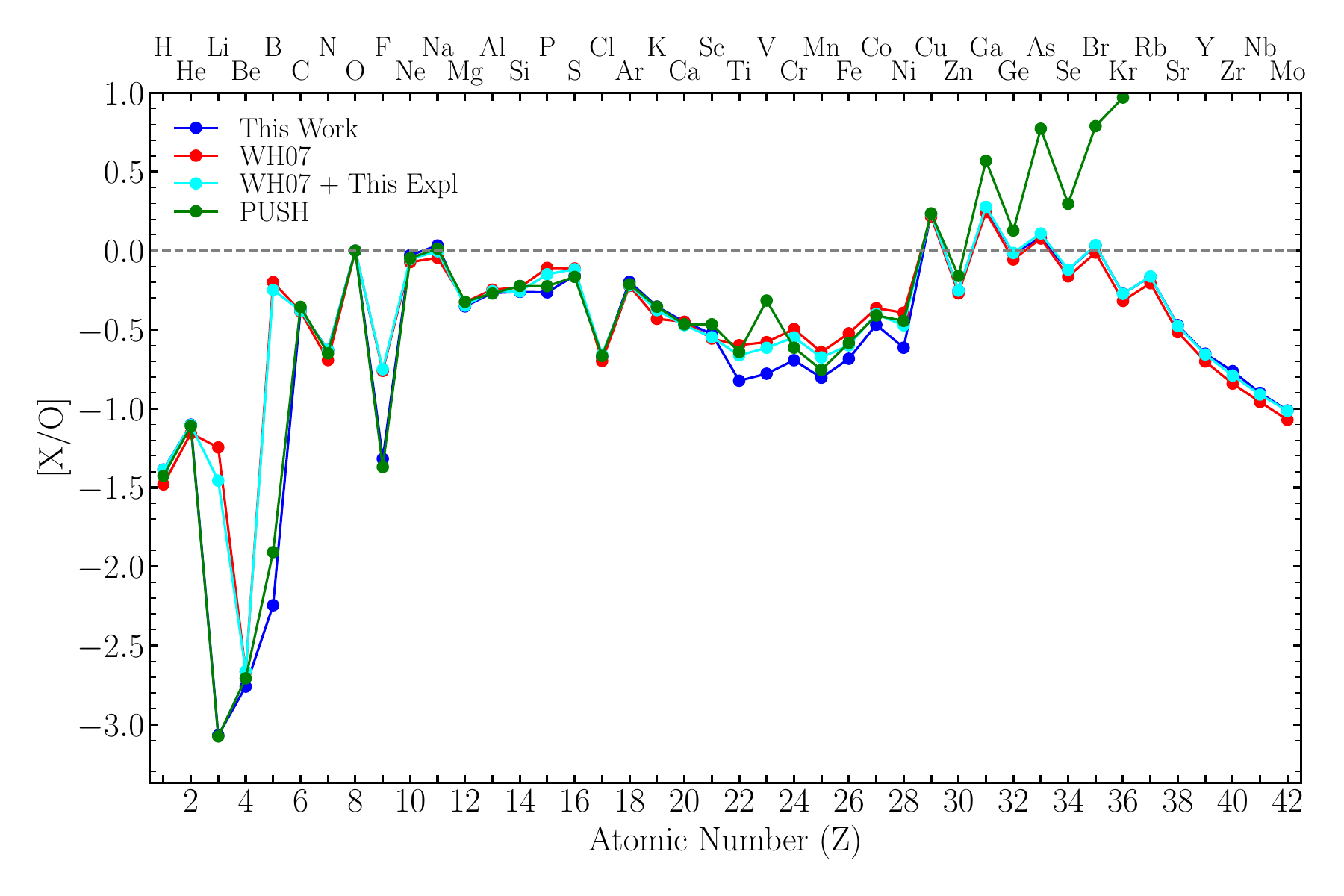}
    \caption{Production factors $\rm [X/O]$ of element $\rm X$ and oxygen with respect to their solar ratio. We use the solar abundances from \citet{Asplund2009_SolAbu}. A value of zero indicates the same ratio as in the solar system. The yields of each star are weighted by a Salpeter IMF and then normalized to one. The blue line shows the yields obtained in this work; the red line shows the yields from WH07 assuming that everything explodes; the cyan line shows the yields from WH07 adopting the explodability derived in this work; and the green line shows the yields from C19.}
    \label{fig:X_over_O_WH07}
\end{figure}

When we compare our simulations with WH07 (where everything explodes), the main difference is that less hydrogen is injected overall into the ISM. This is because in our simulations the 12, 13, 14, 15, 18, 28, and 100 $M_\odot$ stars do not explode and therefore contribute to the ISM only through their winds. In the WH07 set, however, these stars eject their H envelopes; therefore, the abundances of every other element are comparatively reduced. This explains the larger yields of C, O, Ne, Na, Cl, Ar, and K in our simulations. Similarly, the trans-Fe elements also comparatively increase, since their abundance roughly scales with hydrogen, given that they mostly originate from the initial composition of the star and are not reprocessed during the stellar evolution or explosion. Therefore, imposing our explodability on the WH07 set significantly improves the agreement for all these elements. However, the discrepancy for Fe-peak elements remains. We thus conclude that this discrepancy is not due to explodability but rather to different explosion dynamics. As further confirmation, the production factors in Fig.~\ref{fig:X_over_O_WH07} show that normalizing the yields to oxygen significantly reduces the discrepancy caused by explodability for all elements except the Fe-peak ones.

Since the amount of material synthesized during pre-supernova and explosive nucleosynthesis scales with O, it is not surprising that the explodability-related discrepancy is smaller. It should, however, be noted that the ratio plotted in Fig.~\ref{fig:Element_comparison_WH07} is linear, whereas $\rm [X/O]$ is logarithmic and therefore naturally hides small discrepancies. Nonetheless, it is quite clear that the discrepancy in light elements disappears, while the one in Fe-peak elements remains -- even when imposing our explodability on the WH07 yields -- further confirming that the different explosion dynamics are responsible.

As outlined in Sect.~\ref{sec:expl_dyn}, WH07 fixes the explosion energy at 1.4~B for all of the explosions, and as shown in Fig.~\ref{fig:expl_ene_obs}, most of our simulations have lower explosion energies. This explains the smaller amounts of Fe-peak elements, which directly correlate with the amount of ejected $\ce{^56Ni}$.

Lastly, WH07 find much larger abundances of fluorine and phosphorus since they include the effects of $\nu$-process \citep{Woosley1990_nu_proc}, responsible for synthesizing several rare isotopes, such as $\ce{^19F}$ and $\ce{^31P}$, through spallation reactions. Given the absence of neutrinos from our calculations, and the fact that neutrino spallation reactions are not included in \textsc{SkyNet}, we do not discuss the nucleosynthesis of these peculiar isotopes.

\subsubsection{PUSH}
We also compared our yields with those from C19, who also employed a neutrino-driven explosion from the PUSH code \citep{Perego2015_PUSH1} in their calculations. Figure~\ref{fig:Element_comparison_WH07} shows that our yields are overall larger for everything except Ti, V, Ni, and all the trans-Fe elements. For the first three elements, the isotopes causing the discrepancies are $\ce{^47Ti}$, $\ce{^48Ti}$, and $\ce{^51V}$ and $\ce{^58Ni}$. This can be partially explained by the contribution from the $\nu p$-process, which is not included in this work. Additionally, part of the discrepancy for these isotopes, in particular for $\ce{^47Ti}$ and $\ce{^51V}$, might be caused by the weak r-process in the C19 models, which also accounts for their overproduction of trans-Fe elements. However, given the more approximate neutrino transport adopted in PUSH, it is not clear how much of the neutron-rich material in the one or two zones closest to the mass cut \citep{Ghosh2022_PUSH_EOS} is a numerical artifact (Curtis, Perego, private communication). Part of the reason for the larger yields of $\ce{^58Ni}$ and $\ce{^48Ti}$ can, however, be attributed to the higher NSE threshold used in C19, which can also affect other key isotopes such as $\ce{^44Ti}$ and $\ce{^57Ni}$, as discussed later in Sect.~\ref{sec:radioactives}.

The larger yields for all other elements then purely result from the different explodabilities shown in Fig.~\ref{fig:explodability_WH07}. If one compares each progenitor separately, the agreement is indeed remarkably good (see Appendix~\ref{app:all_progs_WH07} for a detailed model-by-model comparison). The main difference is that in our models, the lower-mass stars ($\sim 12-15 M_\odot$) do not explode, whereas the higher-mass stars ($\sim 22-25 M_\odot$) do. Therefore, in our models, all the elements from carbon to iron are enhanced, whereas hydrogen and helium are lower, since none of our $12-15 M_\odot$ stars explode and thus eject no hydrogen in the ISM. 

Conversely, the $22-25~M_\odot$ stars explode -- contrary to the assumption in C19 -- and since they have very large CO cores, they comparatively contribute much more to the overall ISM abundance, explaining the larger yields of carbon, oxygen, and other lighter elements. This is further confirmed by normalizing the yields to oxygen (as shown in Fig.~\ref{fig:Element_comparison_WH07}), which improves agreement except, of course, for trans-Fe elements, vanadium, titanium, and nickel.

\subsection{LC18}
\label{sec:comparison_LC18}
As mentioned above, LC18 use a bomb model launched with the \verb|HYPERION| code \citep{Limongi2020_HYPERION} to simulate shock propagation in their explosive nucleosynthesis calculations. In their work, \citet{LC18} launched a shock wave from inside the iron core (at $\approx 1 M_\odot$), tuning the explosion energy so that the entire envelope could be unbound. In more realistic SN simulations, such as those discussed here, neutrinos continuously deposit energy behind the shock during its initial expansion through the iron core. Once the shock breaks out of the iron core, it becomes a weak, standing accretion shock and is re-energized only a few hundred milliseconds later, when the explosion energy begins to grow. Therefore, with a bomb inside the iron core, a huge part of the core must be ejected. This requires a much stronger shock than that seen in a self-consistent CCSN simulation. After launching the shock wave, they manually set the mass cut to eject $0.07 M_\odot$ of $\ce{^56Ni}$ for all stars.

LC18 provide four different sets of yields for each progenitor: (i) all stars explode, or all stars $>25 M_\odot$ fail and eject only their winds; (ii) the entire envelope is ejected, or a mixing and fallback model is used (still fixing the ejection of $0.07 M_\odot$ of $\ce{^56Ni}$). We compared our results to the set where everything explodes with no mixing and fallback (their Set F\footnote{From the ORFEO repository: \url{http://orfeo.iaps.inaf.it}}). We also compared with the set where everything explodes with mixing and fallback (set M) but qualitatively found no significant differences since -- as discussed in the remainder of this section -- the mass cut and explosion energy have much greater impact on the overall yields. We decided not to compare with the set where all stars $>25 M_\odot$ fail, since we already account for the much more realistic explodability derived from \textsc{GR1D}.

\begin{figure}
    \centering
    \includegraphics[width=\linewidth]{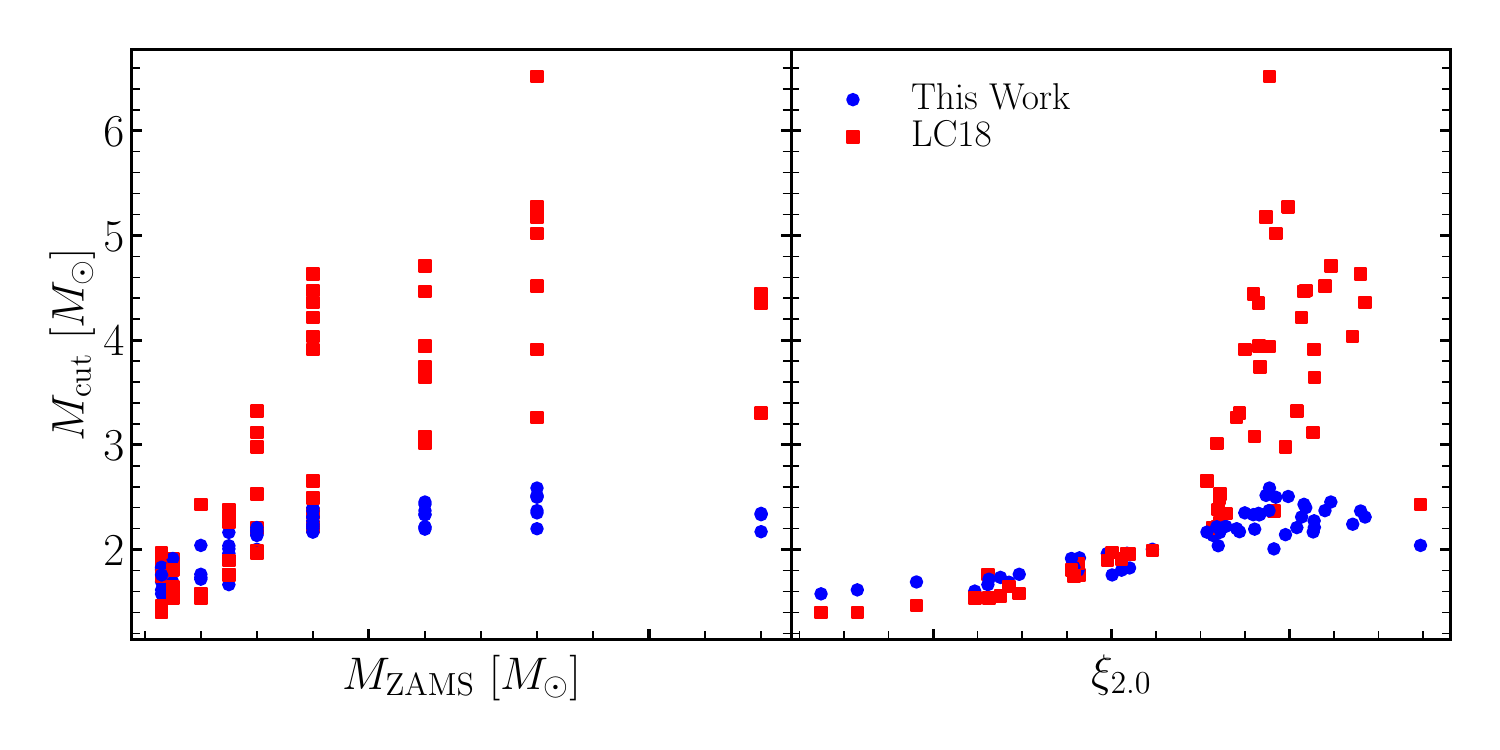}
    \caption{Mass cut for the LC18 progenitor models. The red squares represent the original explosion models of LC18, and the blue circles represent the explosions carried out in this work. Left: Mass cut as a function of initial ZAMS mass. Right: Mass cut as a function of pre-supernova compactness. Note that all metallicities and initial rotational velocities are shown on the same plot. The mass cuts are generally larger for the LC18 explosions, especially at high compactness, although both in this work and in the original LC18 one can see a clear correlation between compactness and mass cut (see the discussion in the text for more details).}
    \label{fig:mcut_LC18}
\end{figure}

The location of the mass cut is plotted in Fig.~\ref{fig:mcut_LC18} for all rotations and metallicities as a function of both the initial ZAMS mass and the pre-supernova compactness of the progenitor (defined in Eq.~\ref{eq:compactness}). As expected, the differences in mass cut (as well as in the mass of the resulting neutron star) are quite large (see Fig.~\ref{fig:mcut_LC18}). This is a consequence of both launching the shock wave from inside the Fe core and ejecting a fixed amount of $\ce{^56Ni}$ for all stars.

Comparing our yields with those of LC18 is slightly more complicated compared to the WH07 case. This is because some of the progenitors rotate and in principle contribute differently to the IMF. To perform a global analysis, we weighted the yields of each star -- for for each metallicity -- by the IMF and by an initial distribution of rotational velocities (IDORV; \citealt{Prantzos2018_GCE_rotation_I}). Following \citet{Rizzuti2021_SrBa_IDORV}, we assumed that the rotational velocities were sampled from a Gaussian distribution with mean
\begin{equation}
    \mu = 
    \begin{cases}
        300 \cdot 0.405 \cdot \exp{\left\{-2.324 \cdot ([{\rm Fe/H}] + 3)\right\}} \ {\rm km/s} \\
        \qquad \qquad \qquad \text{for } [{\rm Fe/H}] \geq -3 \\
        300 \cdot 0.405 \ {\rm km/s} \\
        \qquad \qquad \qquad \text{for } [{\rm Fe/H}] < -3
    \end{cases},
\end{equation}
and standard deviation
\begin{equation}
    \sigma =
    \begin{cases}
    114.2 -58.5 \cdot ([{\rm Fe/H}] + 3) & \text{for } -3 \leq [{\rm Fe/H}] \leq -1 \\
    114.2 & \text{for } [{\rm Fe/H}] < -3 \\
    0 & \text{for } [{\rm Fe/H}] \geq -1
    \end{cases}.
    \label{eq:mu}
\end{equation}

\begin{figure*}
    \centering
    \includegraphics[width=\linewidth]{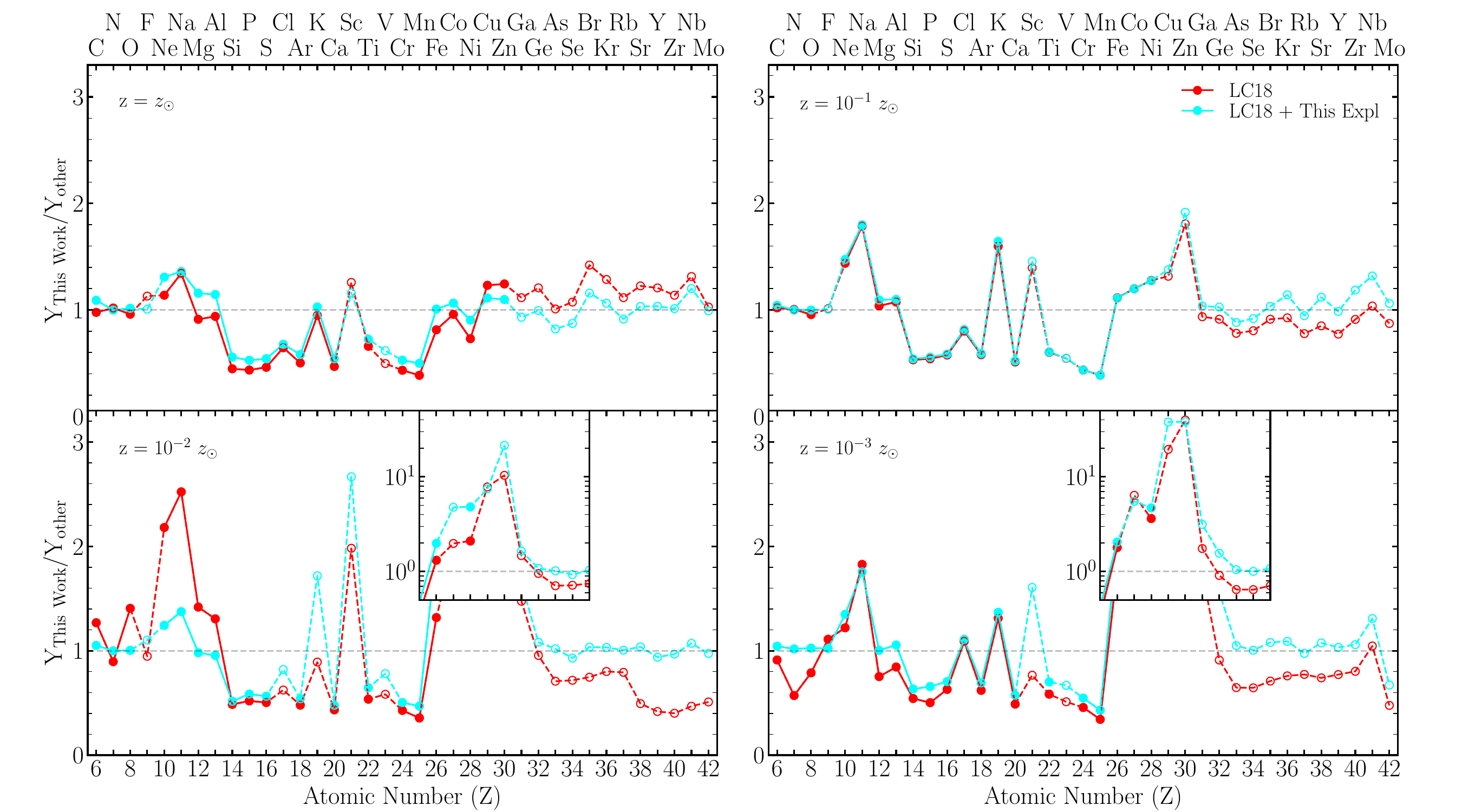}
    \caption{Ratio of the yields obtained in this work to those obtained by LC18, each weighted by a Salpeter IMF, averaged over the IDORV, and subsequently normalized to one. The red line shows the ratio of our yields to the original yields from LC18, for which every star explodes. The cyan line is the same as the red, except that we apply our explodability to the yields of LC18, analogously to what is done in Fig. \ref{fig:Element_comparison_WH07}. The empty circles and the dashed line connecting them indicate elements for which the total yield is less than $10^{-7} M_\odot$. The very large discrepancies at low metallicities are due to the fact that these are typically higher-compactness stars and therefore are expected to eject high amounts of $\ce{^56Ni}$, which is, however, kept fixed at $0.07 M_\odot$ in LC18 (see the text for more details).}
    \label{fig:Element_comparison_LC18}
\end{figure*}

The ratio between our yields and the original yields from LC18 is shown in Fig.~\ref{fig:Element_comparison_LC18}. As done for the WH07 comparison, we impose our explodability on the LC18 yields, represented by the cyan line. As metallicity decreases, discrepancies, particularly in the Fe-peak region, significantly increase.

\begin{figure*}
    \centering
    \includegraphics[width=\linewidth]{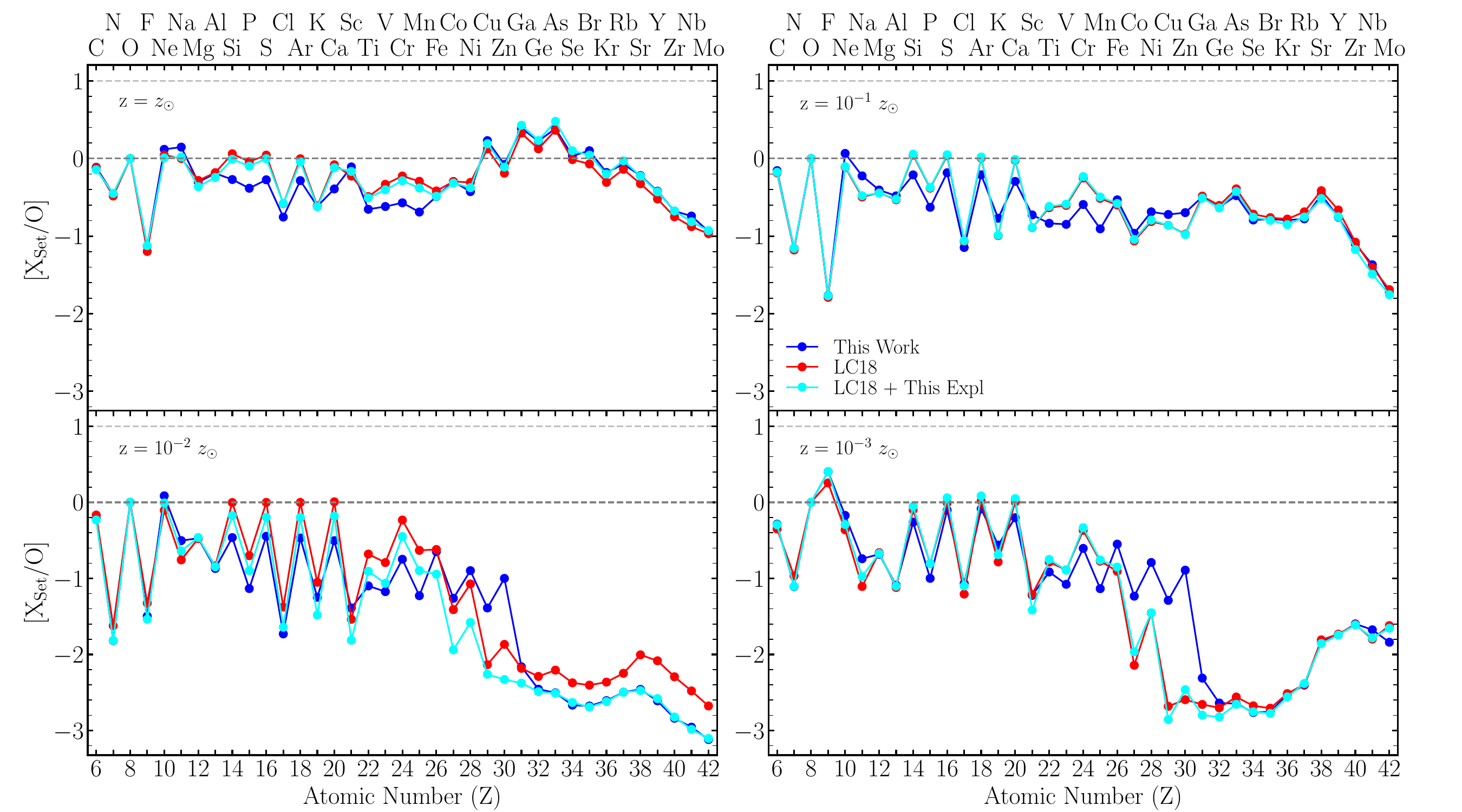}
    \caption{Production factors $\rm [X/O]$ of element $\rm X$ and oxygen with respect to their solar ratio. We use the solar abundances from \citet{Asplund2009_SolAbu}. A value of zero indicates the same ratio found in the solar system. The yields of each star are weighted by a Salpeter IMF, averaged over the IDORV, and then normalized to one. The blue line shows the yields obtained in this work; the red line shows the yields from LC18 assuming that everything explodes; and the cyan line shows the yields from LC18 adopting the explodability derived in this work.}
    \label{fig:X_over_O_LC18}
\end{figure*}

As seen in Fig.~\ref{fig:expl_ene_obs}, the amount of $\ce{^56Ni}$ ejected can vary by an order of magnitude depending on the explosion energy and compactness of the progenitor star. However, as mentioned above, LC18 fixed this amount at $0.07 M_\odot$, potentially introducing large errors in the final abundances of Fe-peak elements. This is because $\ce{^56Ni}$ decays to $\ce{^56Fe}$, the most abundant Fe isotope. Therefore, fixing the amount of $\ce{^56Ni}$ in practice fixes the amount of ejected Fe (and, consequently, all the Fe-peak elements) by the supernova.

For solar metallicity progenitors, $0.07 M_\odot$ represents a reasonable amount of ejected $\ce{^56Ni}$ overall -- quite similar to the the IMF- and IDORV-averaged values from our simulations. At lower metallicity, however, stars are typically more compact; therefore, progenitors with $M_{\rm ZAMS} > 30 M_\odot$ are expected to eject very large amounts of $\ce{^56Ni}$, as shown by the most energetic explosions in Fig.~\ref{fig:expl_ene_obs}. An important caveat is that some of these stars are expected to die as BHSNe and thus will likely experience significant fallback that severely decrease both their explosion energy and ejected $\ce{^56Ni}$ mass \citep{Eggenberger2025_BHSNe_EOS}. Therefore, more accurate models are needed to investigate the full impact of these high-mass stars on the overall abundance of Fe-peak elements. We leave this to future work. Nonetheless, we do not expect our conclusions to qualitatively change, even when considering the uncertainties in BHSNe explodability and ejecta mass. As shown in detail in Appendix~\ref{app:all_progs_LC18} and as shown in Fig.~\ref{fig:expl_ene_obs}, only five progenitors eject less than $0.07 M_\odot$ of $\ce{^56Ni}$ in our simulations. Therefore, we can robustly conclude that at low metallicity, we expect much higher abundances of Fe. 

Within the Fe-peak, we find significant differences between the elements to the left (Ti, V, Cr, and Mn) and the right (Co, Ni, Cu, and Zn) of iron. In this work, we obtain significantly more Co, Ni, Cu, and Zn compared to LC18, but less Ti, V, Cr, and Mn. However, as with Fe, different explodability is not the main cause of this discrepancy, as shown by the cyan and red lines in Fig.~\ref{fig:Element_comparison_LC18}. What determines these discrepancies boils down to the different peak temperatures of the ejecta. By starting the explosion inside the iron core, LC18 produce explosion energies that can be quite high, causing large mass cuts (see Fig.~\ref{fig:mcut_LC18}). In general, the higher the explosion energy, the more $\ce{^56Ni}$ can be produced during incomplete Si burning. This is because higher-energy explosions have peak temperatures that decrease much more slowly with mass. Thus, especially for high-compactness progenitors, this allows larger regions of the star to reach typical incomplete Si-burning temperatures (around $4-5$~GK) in LC18, compared to our simulations where the peak temperatures decrease much faster. These regions produce just enough $\ce{^56Ni}$ to reach the $0.07 M_\odot$ threshold, resulting in very large mass cuts in LC18 at peak temperatures of around $\sim 5$~GK. Consequently, contrary to our simulations, no material reaching NSE is ejected in LC18, explaining the larger Co, Ni, Cu, and Zn yields in our models. For the same reason -- slower peak temperature decline -- the region of the star experiencing incomplete silicon burning is larger in LC18, explaining our lower Ti, V, Cr, and Mn yields. This is further evidenced by the lower abundances of elements from Si to Ar in our simulations (see Figure~\ref{fig:Element_comparison_LC18}).

To summarize, LC18 models have very narrow (or completely absent) zones of complete Si burning, and large zones of incomplete Si burning due to their less steep radial dependence of $T_{\rm peak}$ compared to our simulations. This translates into more Co, Ni, Cu, and Zn and less Si, P, S, Ti, V, Cr, and Mn in our models. The caveat mentioned above should, however, be reiterated: our models (and, to be precise, also the presently adopted set F from LC18) ignore the fallback for high-compactness progenitors, which could significantly affect the complete and incomplete Si-burning zones. That said, we find that the above discussion on Fe-peak elements also applies to lower-compactness progenitors (see Appendix~\ref{app:all_progs_LC18} for a star-by-star comparison); therefore, we do not expect these conclusions to qualitatively change with fallback included.

Given the strong impact of explosion energies and mass cuts on the final yields, explodability is of secondary importance here. The main effects are on trans-Fe elements, as also found in the comparison with WH07. Since most of these elements are in the outer layers, the difference in mass cut does not significantly affect them. Interestingly, rotating models at low metallicity (with significant $s-$process nucleosynthesis) and high compactness (with the strongest explosion) show significant destruction of some of these elements during the LC18 explosion (producing other trans-Fe elements). However, this represents at most a $30-40 \%$ difference (e.g., Nb and Sr), which is not very significant for these $s-$process elements.

\subsection{F23}
\label{sec:comparison_F23}
The F23 models analyzed in this work are those from \citet{Farmer2021_Carbon_binary}, who performed explosive nucleosynthesis calculations using a bomb model to simulate shock propagation. In particular, they launched a shock wave from the mass zone where the entropy per baryon exceeds 4, tuning the explosion energy at shock breakout to 1 B.

\begin{figure}
    \centering
    \includegraphics[width=\linewidth]{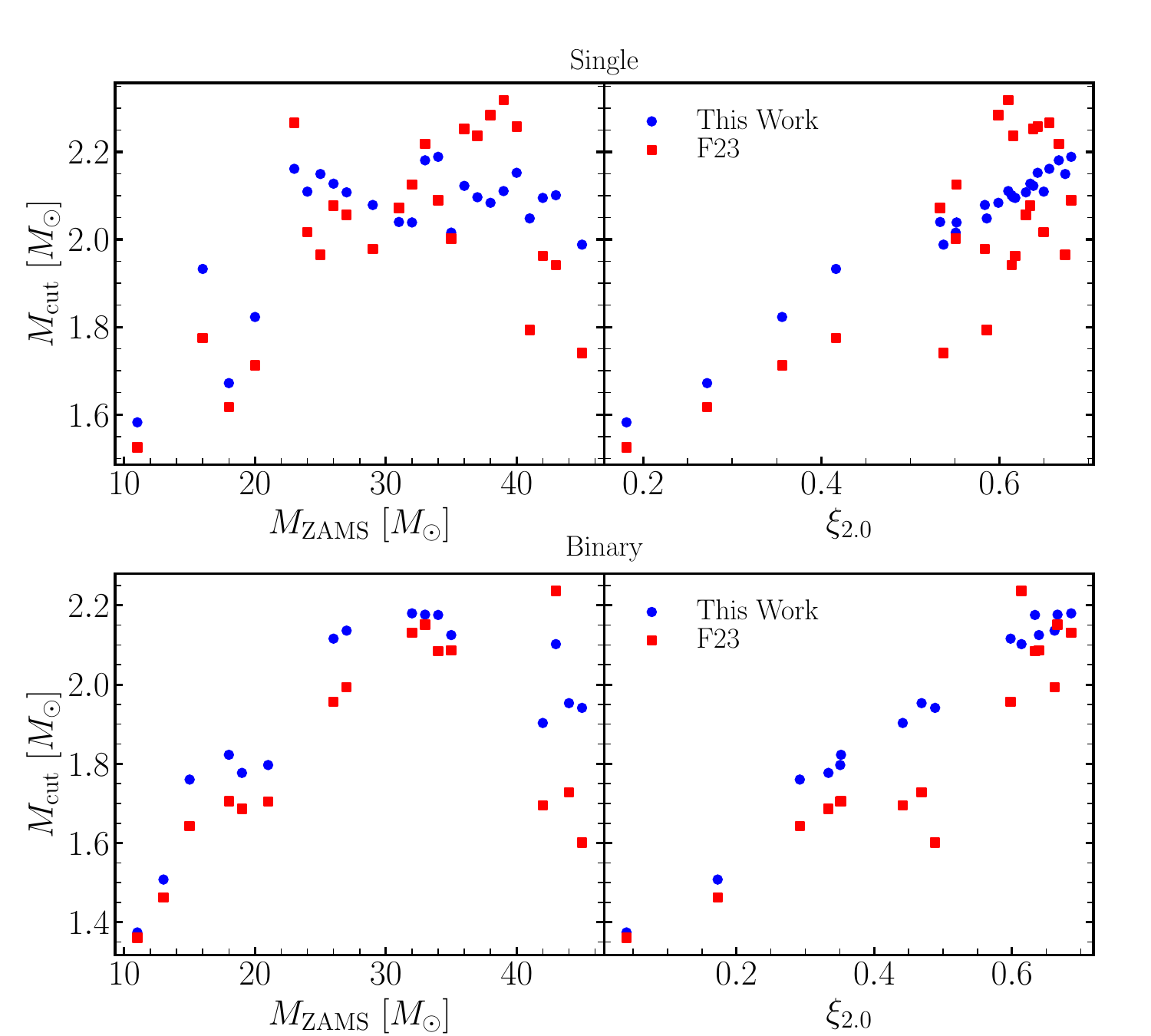}
    \caption{Mass cut for single (top) and binary (bottom) stars for the F23 progenitor models. The mass cut is equal to the baryonic mass of the cold neutron star. The red squares represent the original F23 explosion models, and the blue circles represent the explosions carried out in this work. Left: Quantities shown as a function of initial ZAMS mass. Right: Quantities shown as a function of pre-supernova compactness. The mass cut is larger in the models simulated in this work, and as expected, there is a clear correlation with compactness. However, the scatter in the mass cut for high-compactness progenitors is much larger for the F23 single-star models (see the discussion in the text for more details).}
    \label{fig:mcut_F23}
\end{figure}

The location of the mass cut is plotted for single and binary-stripped stars in Fig.~\ref{fig:mcut_F23} as a function of both initial ZAMS mass and compactness of the pre-supernova progenitor (defined in Eq.~\ref{eq:compactness}). Our models with $\xi_{2.0} < 0.5$ show larger $M_{\rm cut}$, whereas at higher compactness the F23 models show large scatter around our values for both single and binary stars. This can be explained by the fact that for high-compactness progenitors, explosions are rarely triggered by Si/O interface accretion \citep{Boccioli2024_FEC+_SiO}, but by increasingly efficient neutrino heating \citep{Boccioli2025_comp_vs_Qdot}. 

\begin{figure}
    \centering
    \includegraphics[width=\linewidth]{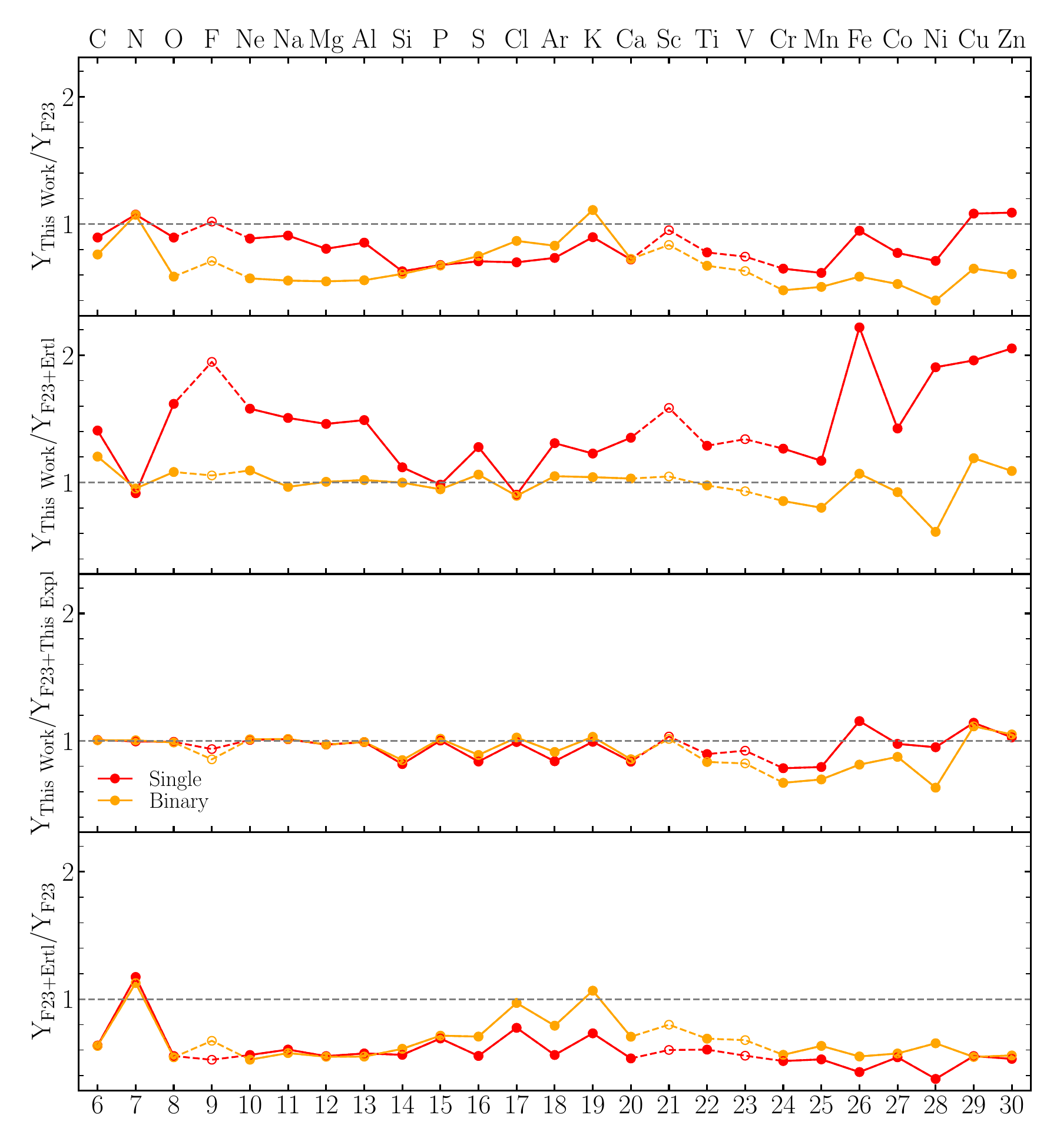}
    \caption{Ratio of the yields obtained in this work to those obtained in previous studies, each weighted by an IMF and subsequently normalized to one. The red lines show the results obtained for single stars, and the orange lines show the results obtained for binary stars. The empty circles and the dashed line connecting them indicate elements for which the total yield is less than $10^{-7} M_\odot$. Top panel: Ratio of our yields to the original yields of F23, assuming that everything explodes. Second panel: Ratio of our yields to the original yields of F23, adopting the explodability from E16. Third panel: Ratio of our yields to the original yields of F23, adopting the explodability found in this work. Bottom panel: Ratio of the F23 yields adopting the explodability from E16 to the F23 yields assuming that everything explodes. }
    \label{fig:Element_comparison_F23}
\end{figure}

\begin{figure}
    \centering
    \includegraphics[width=\linewidth]{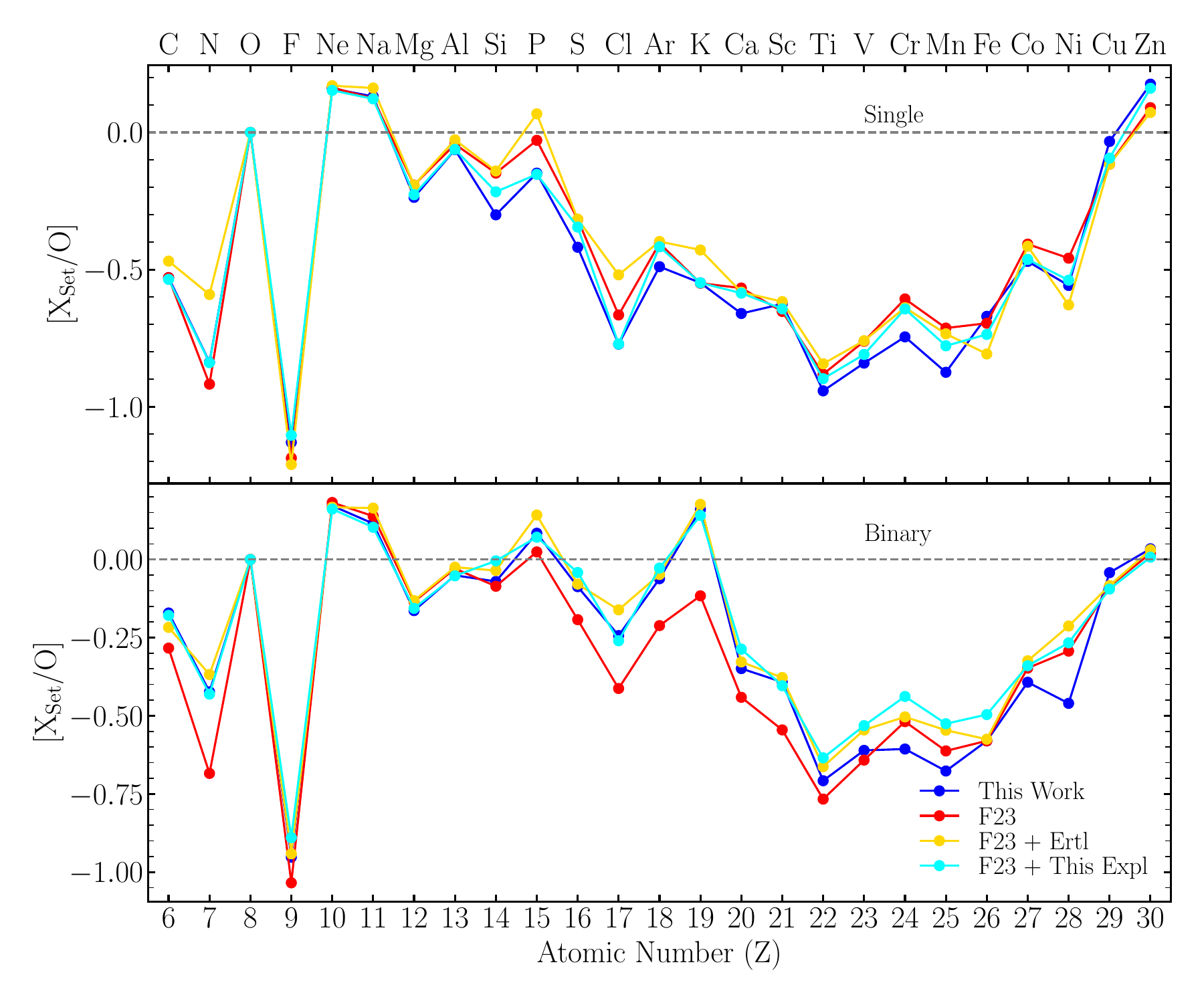}
    \caption{Production factors $\rm [X/O]$ of element $\rm X$ and oxygen with respect to their solar ratio. We use the solar abundances from \citet{Asplund2009_SolAbu}. A value of zero indicates the same ratio found in the solar system. The yields of each star are weighted by a Salpeter IMF and then normalized to one. Top and bottom: Results for single and binary stars, respectively. The blue line shows the yields obtained in this work; the red line shows the yields from F23 assuming that everything explodes; the orange line shows the yields from F23 adopting the explodability from E16; and the cyan line shows the yields from F23 adopting the explodability derived in this work.}
    \label{fig:X_over_O_F23}
\end{figure}

In F23, the authors impose the explodability from E16 on their yields to assess its role. Hence, we also included it in our analysis. As in the discussion above, we compare the overall IMF-weighted yields for single and binary stars separately. Since the IMF for binary-stripped stars is unknown, we assume it is also a Salpeter IMF, as in the case for single stars. In Fig.~\ref{fig:Element_comparison_F23}, we show the ratio between the yields obtained here and those obtained from the F23 explosions. In the top panel, we assume that all models explode for the F23 set; in the second panel, we impose the explodability from E16; in the third panel, we impose the explodability obtained in this work; for completeness, in the bottom panel, we show the ratio between the yields from F23 with the imposed explodability from E16 and those where every star explodes. 

Overall, our yields for elements above carbon are smaller than or equal to F23's, as shown in the upper three panels of Fig.~\ref{fig:Element_comparison_F23}. Let us first analyze the third panel (i.e., our yields vs. the F23 yields with our explodability imposed), where only the different mass cut and explosion dynamics affect the yields. As pointed out in previous sections, different explosion dynamics mostly affect Fe-peak elements, which are less abundant in our explosions. This stems from F23's fixed explosion energy, unlike our more realistic and physically consistent range. F23's $1 \rm B$ value is too high overall, especially for lower-mass stars whose greater IMF weights contribute more to the yields. Thus, our models predict $\sim 70 \%$ less Fe-peak elements in the ISM. Notably, no significant difference appears between single and binary stars. 

Now, let us analyze the top panel of Fig.~\ref{fig:Element_comparison_F23} (i.e., our yields vs. F23's, where every star explodes). As expected, the discrepancy is larger than that shown in the third panel due to different explodability. In particular, the discrepancy in Fe-peak elements is even larger, and there is also a discrepancy in lighter elements. Moreover, the discrepancy in lighter elements differs for single and binary stars due to the explodability of high-mass stars. One can think of the effect of explodability as altering mass weights in a certain range on the overall ISM contribution. In general, fewer low-mass stars from $\sim 12 M_\odot$ to $\sim 20 M_\odot$ explode, injecting less Fe-peak material into the ISM. Similarly, fewer high-mass stars $\gtrsim 20 M_\odot$ explode, injecting less oxygen and nearby light elements into the ISM. From the \textsc{GR1D+} explodability shown in Fig.~\ref{fig:explodability_F23}, most high-mass single stars explode, whereas binary-stripped stars with masses between $37 M_\odot$ and $41 M_\odot$ form an island of failed explosions. This injects much less oxygen and light elements into the ISM, explaining the lower binary-star yields for these elements compared to F23.

A similar argument can be applied to the ratio between our yields and those from F23 with E16 explodability imposed (i.e., Fig.~\ref{fig:Element_comparison_F23}, second panel). For binary stars, the agreement is quite good except for the lower Fe-peak elements in our work, due to F23's fixed explosion energy across all stars and E16's higher low-mass explodability. For single stars, however, our yields are higher overall because, according to E16, most stars (i.e., 22 out of 35) do not explode. Thus, their contribution to ISM is mostly through winds, increasing the relative contribution of hydrogen and helium at the expense of essentially every element above carbon.

For completeness, Fig.~\ref{fig:X_over_O_F23} shows the production factors relative to \citet{Asplund2009_SolAbu} solar abundances for each element compared to oxygen. The same conclusions from our analysis of Fig.~\ref{fig:Element_comparison_F23} hold here. Namely, high-mass stars not exploding decrease the relative contribution of oxygen and neighboring elements to the IMF-weighted yields, since the large CO-cores of these stars are not ejected. This is confirmed by the roughly explodability-insensitive $\rm [X/O]$ from O to Al in both panels of Fig.~\ref{fig:X_over_O_F23}, as these elements, mainly produced during oxygen burning, depend only on oxygen abundance. All the other elements in principle depend on explodability, as proved by the large differences between our $\rm [X/O]$ and F23's with and without E16's explodability imposed. When the explodability effect is removed (i.e., cyan vs. blue lines), Fe-peak elements retain significant discrepancies, indicating sensitivity to how the explosion develops (i.e., how the shock propagates).

\section{Conclusions}
\label{sec:conclusions}
In this paper, we presented nucleosynthetic yields for three different sets of CCSN progenitors. The explosion was carried out using state-of-the-art 1D+ simulations, which are much more sophisticated than those commonly carried out with piston and bomb models. Therefore, we obtained ejected $\ce{^56Ni}$ masses and explosion energies compatible with observations as well as state-of-the-art 3D CCSN simulations. Additionally, mass cuts arise self-consistently during the explosion instead of being artificially imposed, as done in most studies.

We showed that explodability can have a significant impact on nucleosynthesis. Depending on which stars explode and which ones do not, the amount of light versus heavier elements can vary significantly. It is therefore important to consider realistic explosion criteria when performing population studies. Moreover, we showed how piston and bomb models systematically overestimate the amount of Fe-peak elements produced. This stems from the fact, in those studies, all stars typically explode with roughly the same explosion energy, whereas, as we showed in the paper, there is a clear trend of increasing explosion energy with the pre-supernova compactness of the star.

We also showed that 1D+ explosion models have a significant impact on the extent of different explosive burning regions compared to bomb and piston models. This obviously also affects the production of nuclear species produced far from the PNS, such as short- (and long-) lived radioactive nuclei \citep[e.g., $\rm ^{26}Al$, $\rm ^{36}Cl$, $\rm ^{40}K$, $\rm ^{41}Cl$, $\rm ^{60}Fe$; see][and Table \ref{tab:radio_snippet}]{Curtis2019_PUSHIII_nucleosynthesis,Lawson2022_CCSNe_radioactive_nucleosynthesis,Battino2024_Al26_new_rates_impact,Falla2025_Al26_Fe60_rotation} and p-nuclei \citep[n-deficient isotopes of elements heavier than Fe, such as $\rm ^{92}Mo$ or $\rm ^{130}Ba$; see][and references therein]{Roberti2023_GammaProcess1,Roberti2024_z0_expl_nucl}. We aim to extend the discussion on the production of radioactive species in a forthcoming paper.

We showed that when carrying out nucleosynthetic studies in CCSNe, it is important to track the evolution of electron fraction caused by neutrino interactions. Production of selected elements, such as $\ce{^44Ti}$, can have significant contributions from multidimensional effects \citep{Sieverding2023_3D_nucleosynthesis,Wang2024_Ti44_Fornax}. However, there can be significant pre-supernova contributions for selected progenitors (e.g., those that undergo mergers of the convective carbon and oxygen shells a few hours before collapse), which are not usually captured by multidimensional studies given the limited number of simulations that can be run. Moreover, for selected progenitors, 3D simulations predict moderate-to-significant neutron-rich neutrino-driven winds, not found in 1D or 1D+ simulations. Likewise, the amount of proton-rich material (also present in late-time outflows from the PNS) is usually a few times smaller in 1D than 3D simulations \citep{Wang2024_Nucleosynthesis}.

For future studies, it is therefore crucial to consider the uncertainties related to explodability and explosion dynamics (i.e., mass cut and explosion energy). Until large sets of 3D explosion models become available, 1D+ simulations can be a great tool to explore these uncertainties due to their much more robust explosion mechanisms than the old piston and bomb models.

\section*{Data availability}
Yield tables for all nucleosynthesis calculations are available at \url{https://zenodo.org/records/19503168}.

\begin{acknowledgements}
LB is supported by the U.S. Department of Energy under Grant No. DE-SC0004658 and SciDAC grant, DE-SC0024388. LB would like to thank the N3AS center for its hospitality and support, and Dan Kasen, Tianshu Wang, and Sanjana Curtis for fruitful discussions. LR acknowledges the support from the NKFI via K-project 138031, the ChETEC-INFRA -- Transnational Access Projects 22102724-ST and 23103142-ST, and the PRIN URKA Grant Number \verb|prin_2022rjlwhn|. This work was partially supported by the European Union’s Horizon 2020 research and innovation programme (ChETEC-INFRA -- Project no. 101008324), and the IReNA network supported by US National Science Foundation AccelNet (Grant No. OISE-1927130).
\end{acknowledgements}

\bibliographystyle{aa}
\bibliography{MyLibrary.bib}

\begin{appendix}
\onecolumn
\section{Comparison for each progenitor (WH07)}
\label{app:all_progs_WH07}
\begin{figure*}[!ht]
    \centering
    \includegraphics[width=\linewidth]{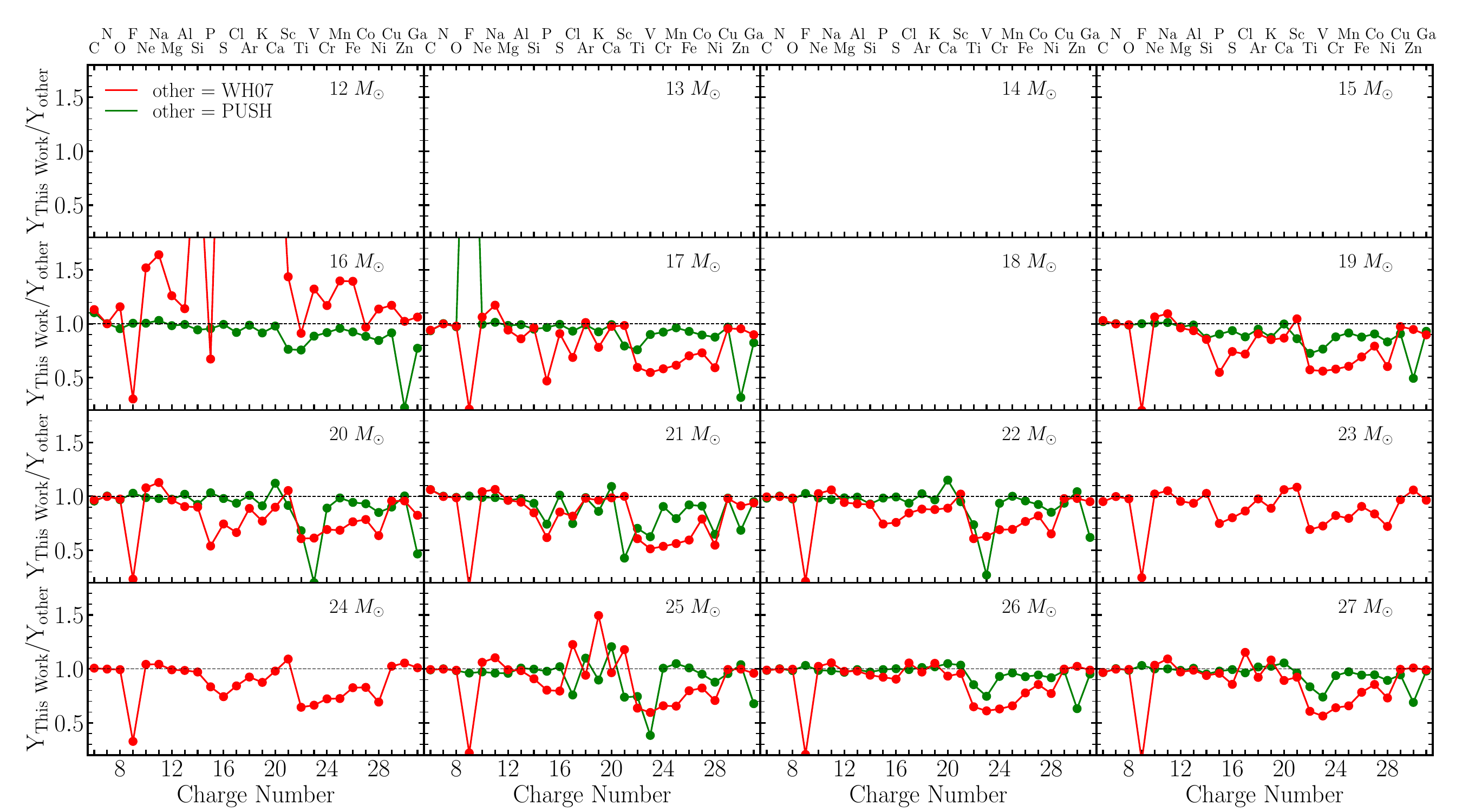}
    \caption{Ratio of our yields to the yields of WH07 (red) and C19 (green) for all progenitors from WH07. Blank panels indicate stars that did not explode in our simulations (see explodability in Figure~\ref{fig:explodability_WH07}). The $16 M_\odot$ progenitor has significantly different yields due to the much larger mass cut reported by WH07, which is an outlier in Fig.~\ref{fig:mcut_WH07}. This also causes the ejecta masses to be different, explaining the large discrepancy seen in the above panel. If C19 predicts that a specific progenitor does not explode, we do not include it in the comparison (e.g., the case of the $23$ and $24 M_\odot$ progenitors.}
    \label{fig:Elements_by_mass_WH07_fig0}
\end{figure*}
\begin{figure*}[!ht]
    \centering
    \includegraphics[width=\linewidth]{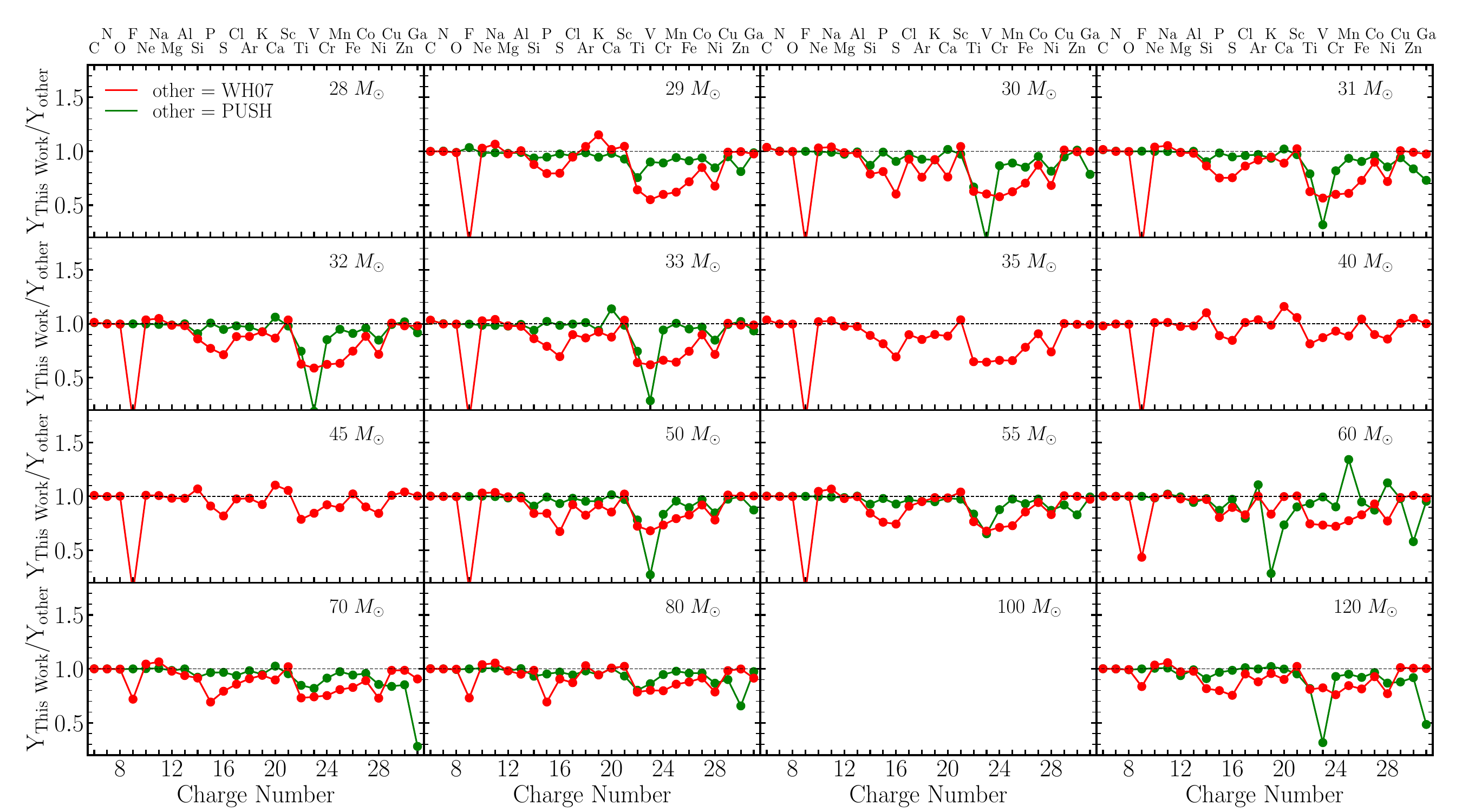}
    \caption{Same as Figure~\ref{fig:Elements_by_mass_WH07_fig0}. }
    \label{fig:Elements_by_mass_WH07_fig1}
\end{figure*}

\newpage
\section{Comparison for each progenitor (LC18)}
\label{app:all_progs_LC18}
\begin{figure*}[!ht]
    \centering
    \includegraphics[width=\linewidth]{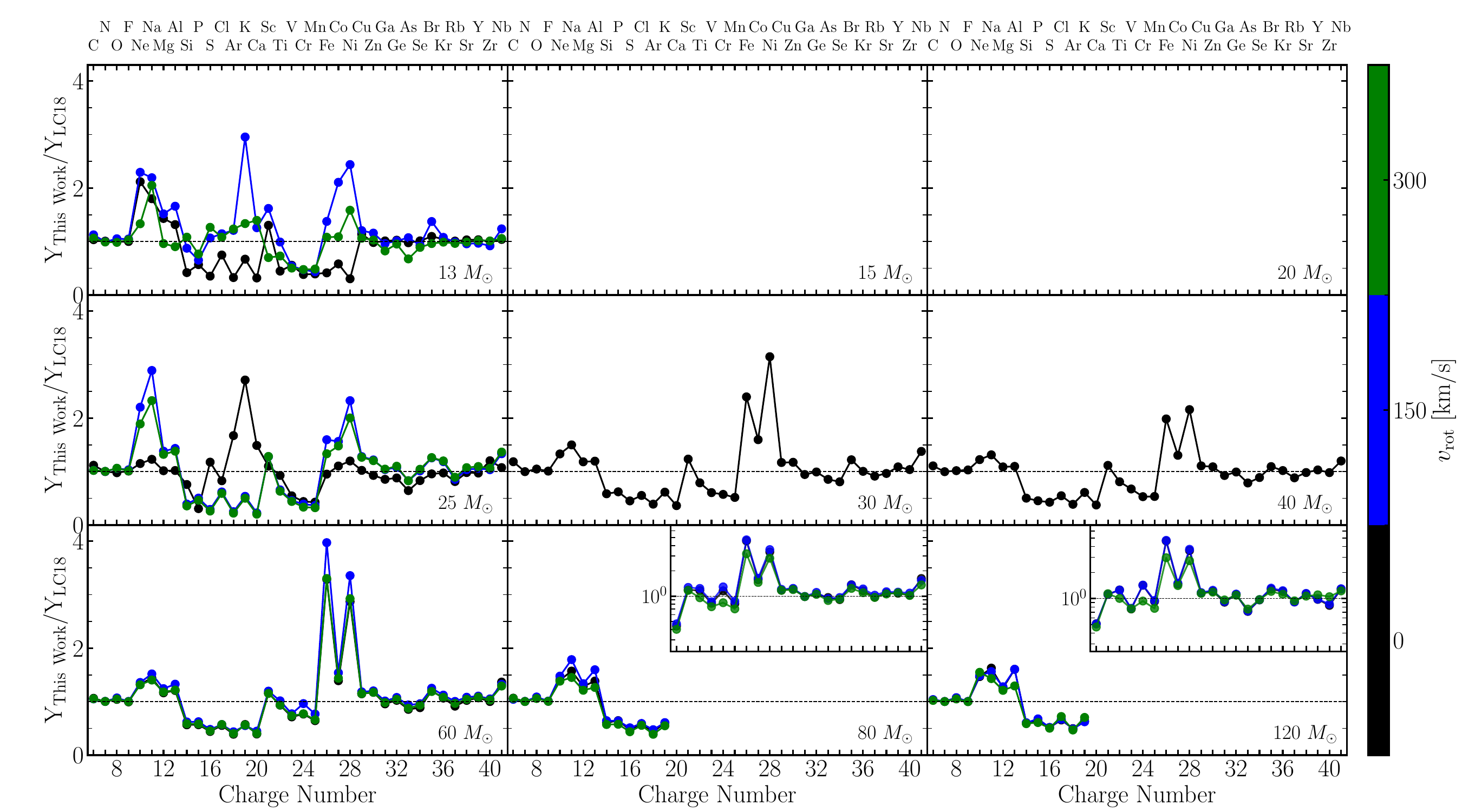}
    \caption{Ratio of our yields to the yields of solar metallicity progenitors from LC18, color coded by initial rotations. If a star did not explode, it is not included in the comparison (e.g., of the $30 M_\odot$ stars at solar metallicity, only the nonrotating one exploded. Blank panels indicate cases where none of the stars with that particular mass exploded.}
    \label{fig:Elements_by_mass_LC18_fig0}
\end{figure*}
\begin{figure*}[!ht]
    \centering
    \includegraphics[width=\linewidth]{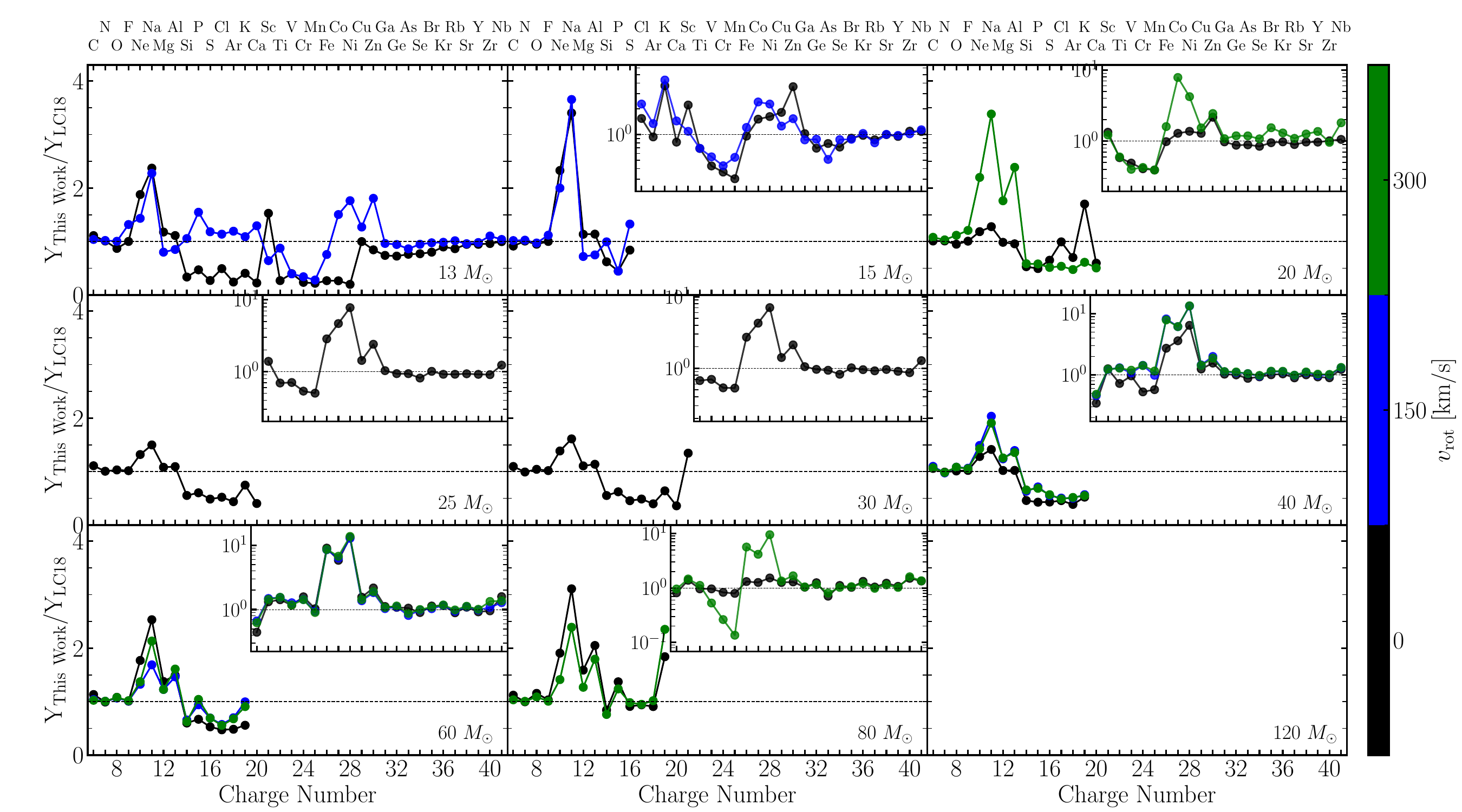}
    \caption{Same as Fig. \ref{fig:Elements_by_mass_LC18_fig0}, but for $\rm [Fe/H] = -1$.}
    \label{fig:Elements_by_mass_LC18_fig1}
\end{figure*}
\begin{figure*}[!ht]
    \centering
    \includegraphics[width=\linewidth]{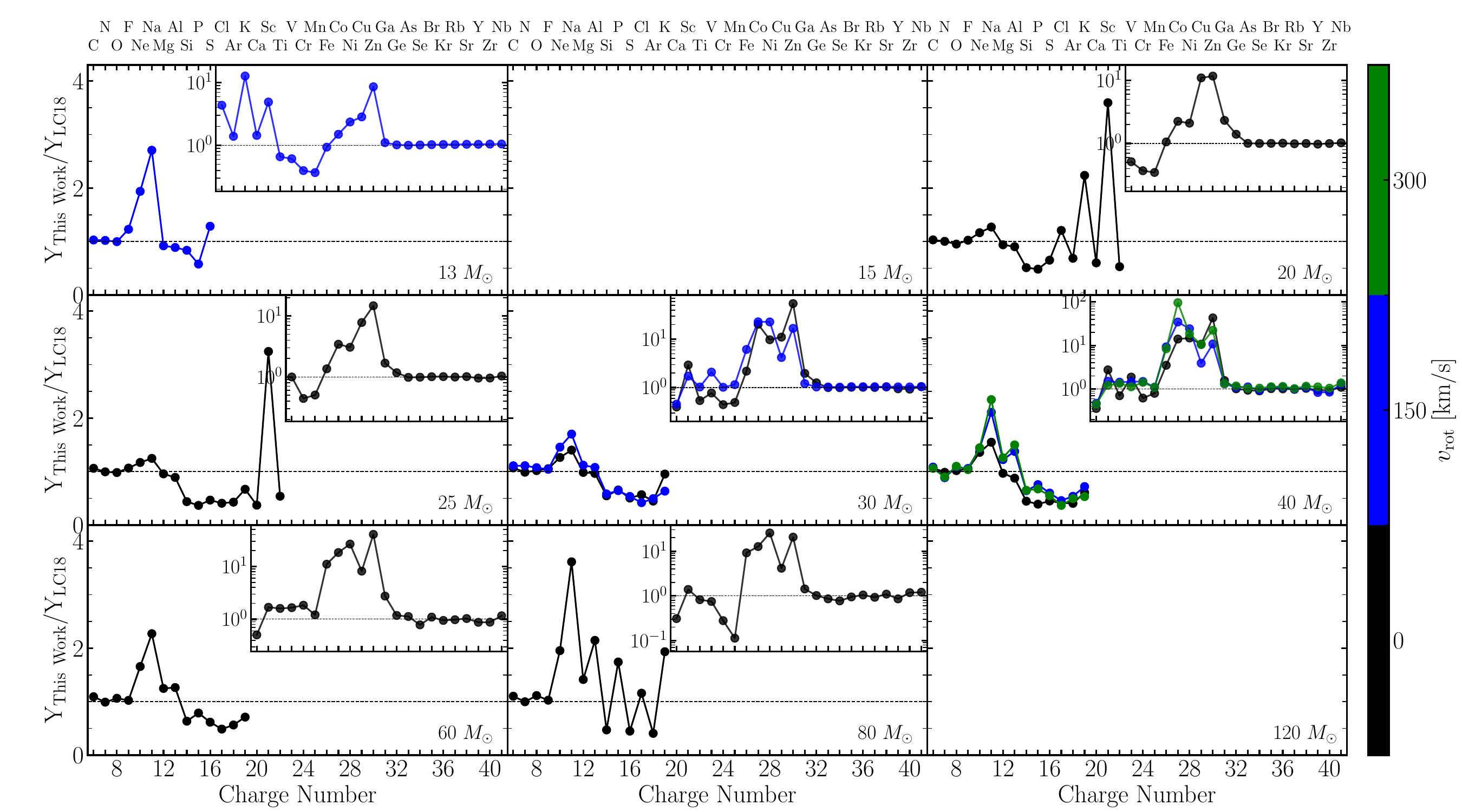}
    \caption{Same as Fig. \ref{fig:Elements_by_mass_LC18_fig0}, but for $\rm [Fe/H] = -2$.}
    \label{fig:Elements_by_mass_LC18_fig2}
\end{figure*}
\begin{figure*}[!ht]
    \centering
    \includegraphics[width=\linewidth]{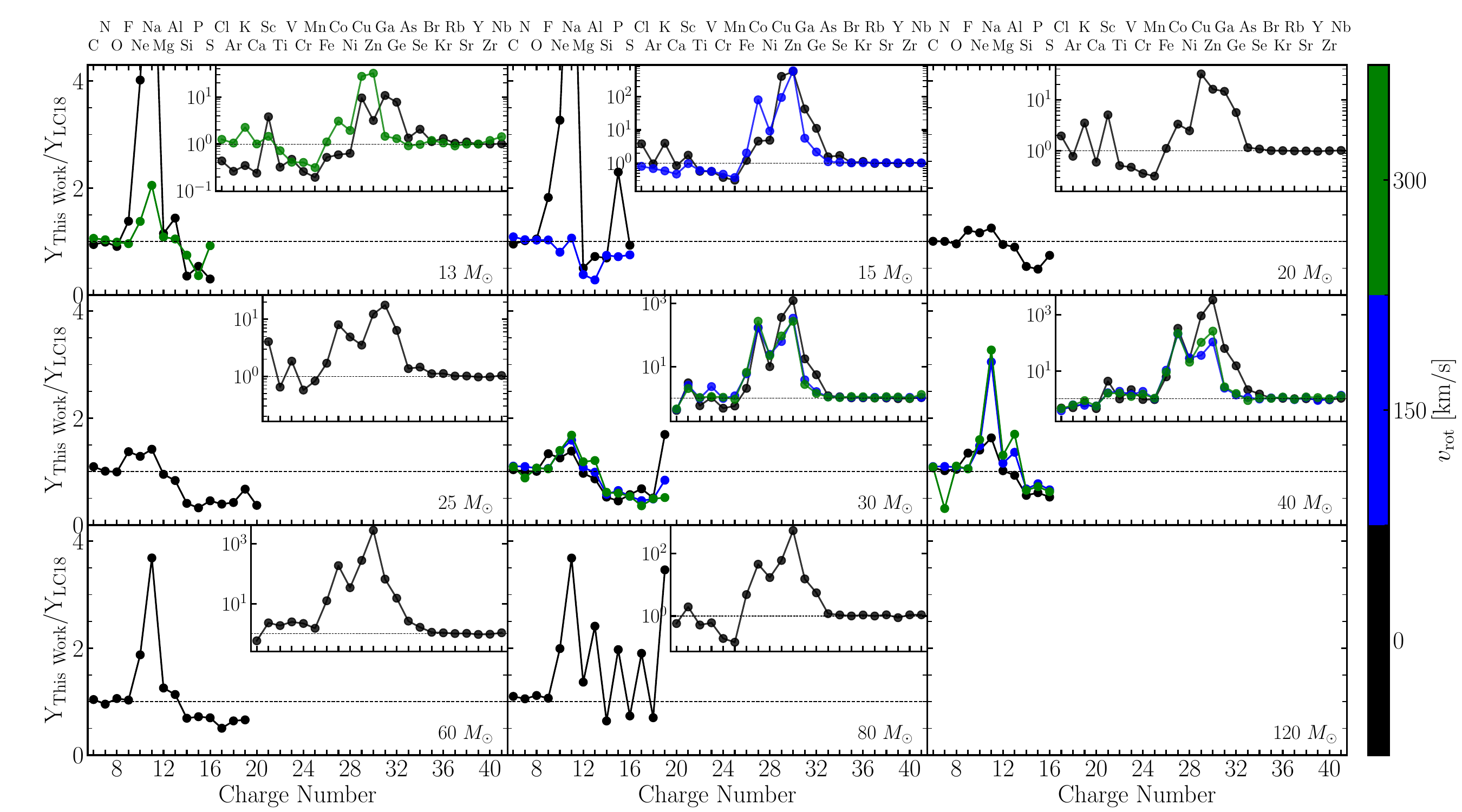}
    \caption{Same as Fig. \ref{fig:Elements_by_mass_LC18_fig0}, but for $\rm [Fe/H] = -3$.}
    \label{fig:Elements_by_mass_LC18_fig3}
\end{figure*}

\clearpage
\section{Comparison for each progenitor (F23)}
\label{app:all_progs_F23}
\begin{figure*}[!ht]
    \centering
    \includegraphics[width=\linewidth]{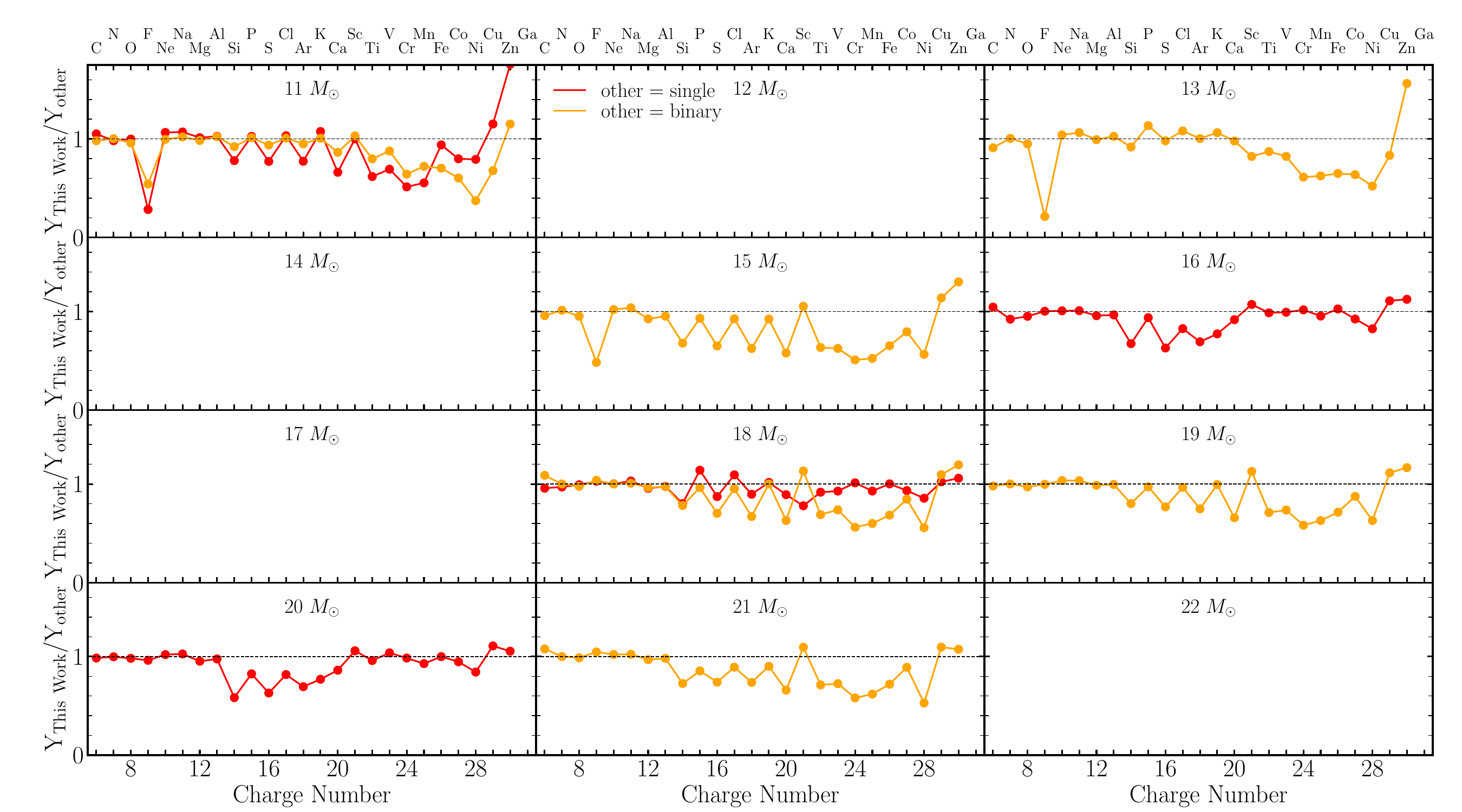}
    \caption{Ratio of our yields to the yields of single-star (red) and binary-stripped (orange) progenitors from F23. If a star did not explode, it is not included in the comparison (e.g., of the $15 M_\odot$ stars, only the binary-stripped one exploded. Blank panels indicate cases where none of the stars with that particular mass exploded.}
    \label{fig:Elements_by_mass_F23_fig0}
\end{figure*}
\begin{figure*}[!ht]
    \centering
    \includegraphics[width=\linewidth]{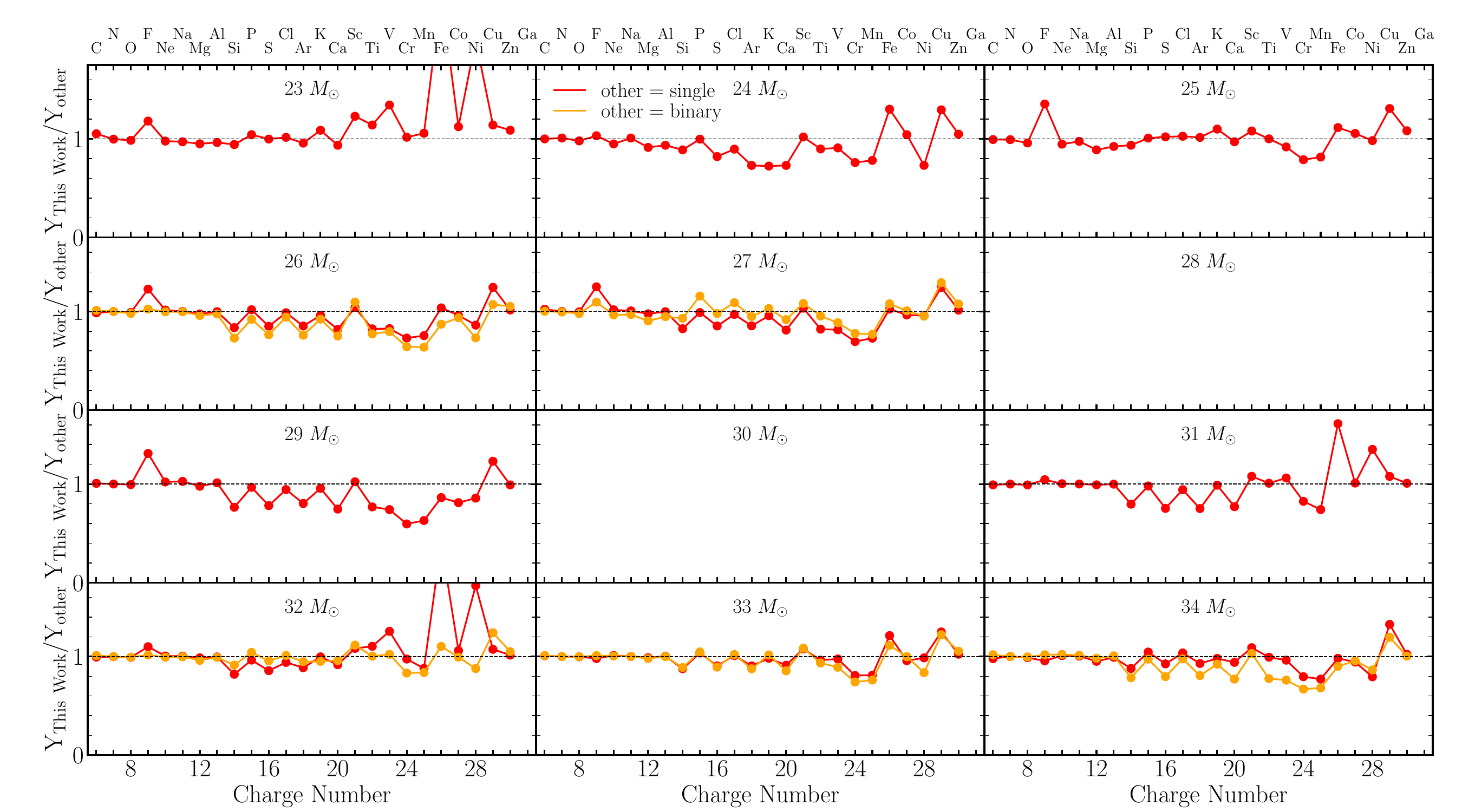}
    \caption{Same as Fig.~\ref{fig:Elements_by_mass_F23_fig0}.}
    \label{fig:Elements_by_mass_F23_fig1}
\end{figure*}
\begin{figure*}[!ht]
    \centering
    \includegraphics[width=\textwidth]{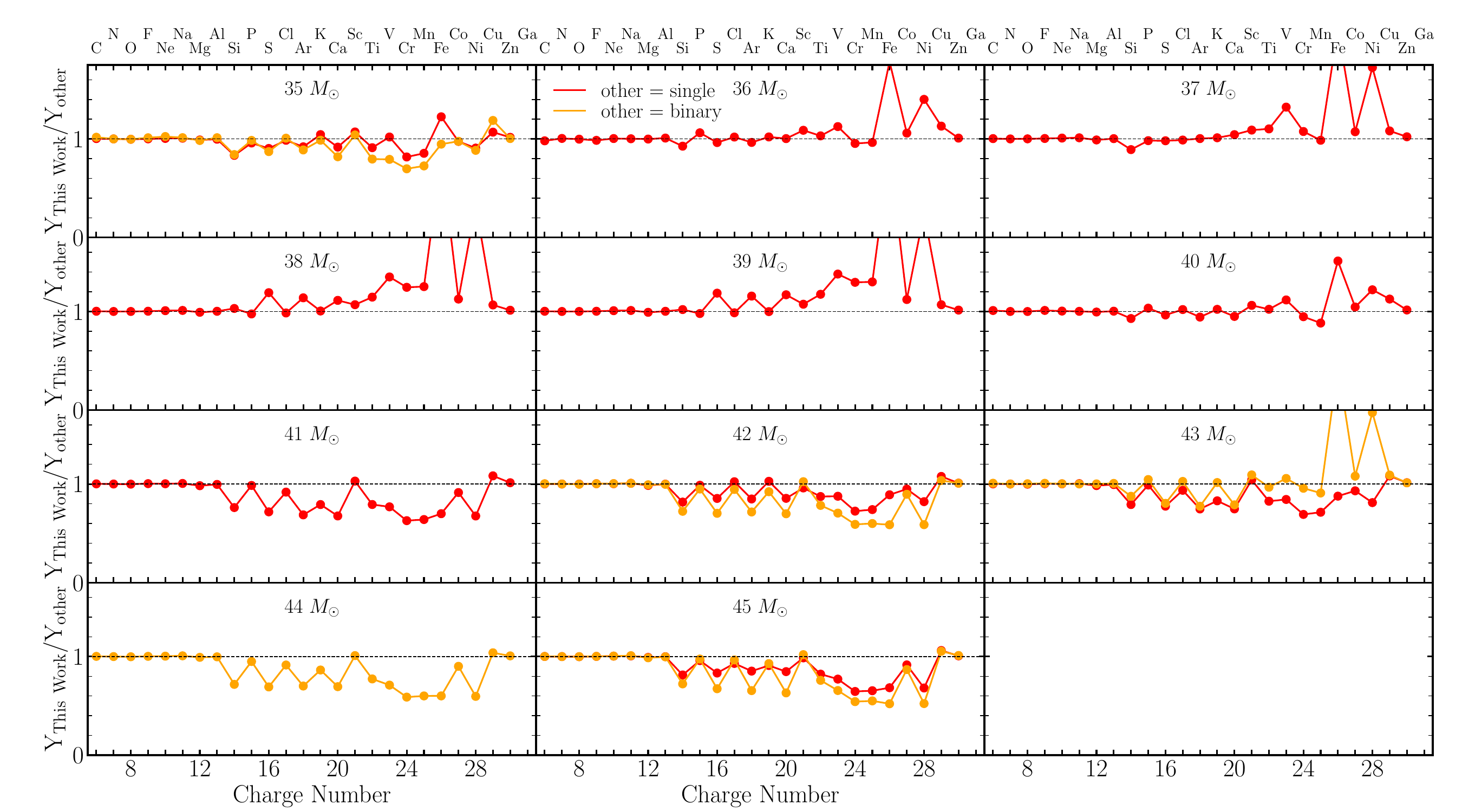}
    \caption{Same as Fig.~\ref{fig:Elements_by_mass_F23_fig0}.}
    \label{fig:Elements_by_mass_F23_fig2}
\end{figure*}

\clearpage

\begin{table}
\section{Yield tables}
\label{sec:appendix_tables}

\centering
\small
\caption{Stable elements for a $25 M_\odot$ single, nonrotating, solar metallicity progenitor.}
\label{tab:ele_snippet}
\begin{tabular}{l|ccc|ccc|ccc}
\toprule
 & \multicolumn{3}{|c|}{WH07} & \multicolumn{3}{|c|}{LC18} & \multicolumn{3}{|c}{F23} \\
 & Post ($M_\odot$) & Wind ($M_\odot$) & $f_{\rm prod}$ & Post ($M_\odot$) & Wind ($M_\odot$) & $f_{\rm prod}$ & Post ($M_\odot$) & Wind ($M_\odot$) & $f_{\rm prod}$ \\
\midrule
C & $3.70 \times 10^{-1}$ & $1.83 \times 10^{-2}$ & 1.04 & $6.94 \times 10^{-1}$ & $2.30 \times 10^{-2}$ & 1.07 & $3.19 \times 10^{-1}$ & $2.03 \times 10^{-2}$ & 1.04 \\
N & $6.36 \times 10^{-2}$ & $1.94 \times 10^{-2}$ & 1.44 & $5.94 \times 10^{-2}$ & $5.53 \times 10^{-2}$ & $1.45 \times 10^{1}$ & $5.22 \times 10^{-2}$ & $3.54 \times 10^{-2}$ & 3.11 \\
O & 3.34 & $5.30 \times 10^{-2}$ & $9.69 \times 10^{-1}$ & 2.43 & $6.73 \times 10^{-2}$ & $9.97 \times 10^{-1}$ & 3.75 & $6.47 \times 10^{-2}$ & $9.52 \times 10^{-1}$ \\
F & $9.09 \times 10^{-6}$ & $3.61 \times 10^{-6}$ & 1.72 & $6.29 \times 10^{-6}$ & $5.72 \times 10^{-6}$ & $1.09 \times 10^{1}$ & $2.44 \times 10^{-5}$ & $3.31 \times 10^{-6}$ & 1.13 \\
Ne & $5.10 \times 10^{-1}$ & $1.15 \times 10^{-2}$ & $9.46 \times 10^{-1}$ & 1.02 & $2.03 \times 10^{-2}$ & 1.01 & $5.89 \times 10^{-1}$ & $1.80 \times 10^{-2}$ & $8.60 \times 10^{-1}$ \\
Na & $7.26 \times 10^{-3}$ & $4.96 \times 10^{-4}$ & $9.50 \times 10^{-1}$ & $2.66 \times 10^{-2}$ & $1.14 \times 10^{-3}$ & 1.01 & $3.94 \times 10^{-3}$ & $7.43 \times 10^{-4}$ & 1.11 \\
Mg & $2.29 \times 10^{-1}$ & $6.65 \times 10^{-3}$ & $9.89 \times 10^{-1}$ & $1.63 \times 10^{-1}$ & $1.17 \times 10^{-2}$ & 1.07 & $3.59 \times 10^{-1}$ & $7.00 \times 10^{-3}$ & $8.14 \times 10^{-1}$ \\
Al & $2.01 \times 10^{-2}$ & $6.10 \times 10^{-4}$ & 1.02 & $1.80 \times 10^{-2}$ & $9.90 \times 10^{-4}$ & 1.04 & $3.35 \times 10^{-2}$ & $6.36 \times 10^{-4}$ & $8.33 \times 10^{-1}$ \\
Si & $2.72 \times 10^{-1}$ & $7.53 \times 10^{-3}$ & $9.09 \times 10^{-1}$ & $1.35 \times 10^{-1}$ & $1.11 \times 10^{-2}$ & $4.52 \times 10^{-1}$ & $5.17 \times 10^{-1}$ & $7.72 \times 10^{-3}$ & $7.47 \times 10^{-1}$ \\
P & $2.72 \times 10^{-3}$ & $6.96 \times 10^{-5}$ & $9.41 \times 10^{-1}$ & $4.74 \times 10^{-4}$ & $9.70 \times 10^{-5}$ & $7.70 \times 10^{-1}$ & $5.67 \times 10^{-3}$ & $8.51 \times 10^{-5}$ & $8.77 \times 10^{-1}$ \\
S & $1.41 \times 10^{-1}$ & $3.83 \times 10^{-3}$ & $7.88 \times 10^{-1}$ & $9.61 \times 10^{-2}$ & $5.15 \times 10^{-3}$ & $5.33 \times 10^{-1}$ & $1.69 \times 10^{-1}$ & $3.85 \times 10^{-3}$ & $5.79 \times 10^{-1}$ \\
Cl & $1.31 \times 10^{-3}$ & $4.98 \times 10^{-5}$ & $7.21 \times 10^{-1}$ & $6.29 \times 10^{-4}$ & $1.37 \times 10^{-4}$ & $7.40 \times 10^{-1}$ & $9.25 \times 10^{-4}$ & $5.11 \times 10^{-5}$ & $4.34 \times 10^{-1}$ \\
Ar & $2.57 \times 10^{-2}$ & $9.98 \times 10^{-4}$ & $9.49 \times 10^{-1}$ & $2.67 \times 10^{-2}$ & $1.11 \times 10^{-3}$ & $5.54 \times 10^{-1}$ & $2.79 \times 10^{-2}$ & $7.60 \times 10^{-4}$ & $4.39 \times 10^{-1}$ \\
K & $1.08 \times 10^{-3}$ & $3.88 \times 10^{-5}$ & $8.31 \times 10^{-1}$ & $6.23 \times 10^{-4}$ & $5.11 \times 10^{-5}$ & $5.73 \times 10^{-1}$ & $3.29 \times 10^{-4}$ & $3.99 \times 10^{-5}$ & $1.69 \times 10^{-1}$ \\
Ca & $1.33 \times 10^{-2}$ & $6.76 \times 10^{-4}$ & 1.69 & $1.86 \times 10^{-2}$ & $1.07 \times 10^{-3}$ & $5.66 \times 10^{-1}$ & $2.06 \times 10^{-2}$ & $6.79 \times 10^{-4}$ & $5.07 \times 10^{-1}$ \\
Sc & $9.81 \times 10^{-6}$ & $4.13 \times 10^{-7}$ & 1.53 & $1.26 \times 10^{-5}$ & $7.74 \times 10^{-7}$ & $4.71 \times 10^{-1}$ & $7.99 \times 10^{-6}$ & $4.29 \times 10^{-7}$ & $2.11 \times 10^{-2}$ \\
Ti & $2.70 \times 10^{-4}$ & $3.12 \times 10^{-5}$ & 2.40 & $2.51 \times 10^{-4}$ & $5.20 \times 10^{-5}$ & $4.17 \times 10^{-3}$ & $4.46 \times 10^{-4}$ & $3.16 \times 10^{-5}$ & $7.86 \times 10^{-3}$ \\
V & $2.83 \times 10^{-5}$ & $3.94 \times 10^{-6}$ & 3.31 & $2.17 \times 10^{-5}$ & $5.29 \times 10^{-6}$ & $7.44 \times 10^{-4}$ & $6.31 \times 10^{-5}$ & $4.03 \times 10^{-6}$ & $9.27 \times 10^{-4}$ \\
Cr & $2.01 \times 10^{-3}$ & $1.80 \times 10^{-4}$ & 7.09 & $1.28 \times 10^{-3}$ & $2.77 \times 10^{-4}$ & $4.81 \times 10^{-3}$ & $5.11 \times 10^{-3}$ & $1.93 \times 10^{-4}$ & $1.28 \times 10^{-2}$ \\
Mn & $9.00 \times 10^{-4}$ & $1.35 \times 10^{-4}$ & 5.50 & $6.58 \times 10^{-4}$ & $1.80 \times 10^{-4}$ & $5.76 \times 10^{-3}$ & $2.34 \times 10^{-3}$ & $1.41 \times 10^{-4}$ & $9.77 \times 10^{-3}$ \\
Fe & $1.56 \times 10^{-1}$ & $1.26 \times 10^{-2}$ & $1.01 \times 10^{1}$ & $1.04 \times 10^{-1}$ & $2.15 \times 10^{-2}$ & $1.51 \times 10^{-1}$ & $5.23 \times 10^{-1}$ & $1.34 \times 10^{-2}$ & $5.98 \times 10^{-1}$ \\
Co & $7.97 \times 10^{-4}$ & $3.68 \times 10^{-5}$ & 1.21 & $9.60 \times 10^{-4}$ & $7.02 \times 10^{-5}$ & $2.01 \times 10^{-2}$ & $1.43 \times 10^{-3}$ & $3.63 \times 10^{-5}$ & $2.42 \times 10^{-2}$ \\
Ni & $1.03 \times 10^{-2}$ & $7.53 \times 10^{-4}$ & 2.80 & $7.73 \times 10^{-3}$ & $1.19 \times 10^{-3}$ & $4.43 \times 10^{-2}$ & $3.36 \times 10^{-2}$ & $7.90 \times 10^{-4}$ & $2.04 \times 10^{-1}$ \\
Cu & $6.80 \times 10^{-4}$ & $8.99 \times 10^{-6}$ & $9.72 \times 10^{-1}$ & $4.54 \times 10^{-4}$ & $1.20 \times 10^{-5}$ & $2.13 \times 10^{-2}$ & $5.26 \times 10^{-4}$ & $9.38 \times 10^{-6}$ & $4.81 \times 10^{-2}$ \\
Zn & $5.41 \times 10^{-4}$ & $2.16 \times 10^{-5}$ & 1.18 & $3.53 \times 10^{-4}$ & $2.89 \times 10^{-5}$ & $2.22 \times 10^{-1}$ & $1.37 \times 10^{-3}$ & $2.32 \times 10^{-5}$ & $9.31 \times 10^{-1}$ \\
\bottomrule
\end{tabular}
\tablefoot{Each column refers to different progenitor sets: WH07, LC18, and F23. Post refers to the post-explosion total ejected yield. Wind refers to the total yield ejected through winds during the pre-collapse evolution. $f_{\rm prod}$ is the production factor calculated as the ratio between the post-explosion and pre-collapse yield (i.e., not including winds). Complete tables in machine-readable form are available online.}
\end{table}

\begin{table}
\centering
\small
\caption{Same as table \ref{tab:ele_snippet}, but for selected long-lived radioactive isotopes.}
\label{tab:radio_snippet}
\begin{tabular}{l|cc|cc|cc}
\toprule
 & \multicolumn{2}{|c|}{WH07} & \multicolumn{2}{|c|}{LC18} & \multicolumn{2}{|c}{F23} \\
 & Post ($M_\odot$) & $f_{\rm prod}$ & Post ($M_\odot$) & $f_{\rm prod}$ & Post ($M_\odot$) & $f_{\rm prod}$ \\
\midrule
$^{26}$Al & $8.77 \times 10^{-5}$ & 1.42 & $1.11 \times 10^{-5}$ & 2.47 & $4.89 \times 10^{-4}$ & $9.31 \times 10^{-1}$ \\
$^{36}$Cl & $7.68 \times 10^{-6}$ & 1.48 & $3.12 \times 10^{-6}$ & $9.19 \times 10^{-1}$ & $2.43 \times 10^{-5}$ & $3.90 \times 10^{-1}$ \\
$^{40}$K & $1.03 \times 10^{-5}$ & 1.69 & $2.37 \times 10^{-6}$ & $9.97 \times 10^{-1}$ & $5.47 \times 10^{-6}$ & $2.59 \times 10^{-1}$ \\
$^{41}$Ca & $5.02 \times 10^{-5}$ & 1.65 & $3.96 \times 10^{-5}$ & $5.98 \times 10^{-1}$ & $1.34 \times 10^{-5}$ & $9.33 \times 10^{-2}$ \\
$^{44}$Ti & $3.97 \times 10^{-5}$ & $2.62 \times 10^{1}$ & $2.15 \times 10^{-5}$ & 2.42 & $8.81 \times 10^{-5}$ & $7.81 \times 10^{-1}$ \\
$^{48}$V & $2.80 \times 10^{-7}$ & 1.75 & $1.63 \times 10^{-7}$ & $2.31 \times 10^{-2}$ & $1.58 \times 10^{-7}$ & $5.16 \times 10^{-4}$ \\
$^{53}$Mn & $1.36 \times 10^{-5}$ & 2.96 & $8.82 \times 10^{-6}$ & $4.93 \times 10^{-4}$ & $3.29 \times 10^{-5}$ & $5.37 \times 10^{-4}$ \\
$^{56}$Ni & $1.19 \times 10^{-1}$ & $5.66 \times 10^{6}$ & $6.97 \times 10^{-2}$ & $4.21 \times 10^{1}$ & $4.70 \times 10^{-1}$ & 5.57 \\
$^{60}$Fe & $1.20 \times 10^{-4}$ & $9.69 \times 10^{-1}$ & $3.76 \times 10^{-5}$ & $1.42 \times 10^{-3}$ & $4.82 \times 10^{-4}$ & $7.87 \times 10^{-2}$ \\
$^{63}$Ni & $4.64 \times 10^{-4}$ & $9.08 \times 10^{-1}$ & $2.76 \times 10^{-4}$ & $2.40 \times 10^{-2}$ & $2.82 \times 10^{-4}$ & $3.17 \times 10^{-2}$ \\
\bottomrule
\end{tabular}
\end{table}

\end{appendix}
\end{document}